\documentclass[galaxies,article,accept,moreauthors,pdftex]{Definitions/mdpi}

\usepackage{comment}

\firstpage{1} 
\pubvolume{1}
\issuenum{1}
\articlenumber{0}
\pubyear{2023}
\copyrightyear{2023}
\externaleditor{Academic Editor: Yosuke Mizuno 
}
\datereceived{25 December 2022} 
\daterevised{16 February 2023} 
\dateaccepted{3 March 2023} 
\datepublished{ } 
\hreflink{https://doi.org/} 

\newcommand\sgra{Sgr\,A$^{*}$\,}
\newcommand{\xc}[1]{\textcolor{black}{#1}}  
\usepackage{gensymb}

\Title{Monitoring the Size and Flux Density of \sgra during the Active State in 2019 with East Asian VLBI Network}

\TitleCitation{Monitoring the Size and Flux Density of \sgra during the Active State in 2019 with East Asian VLBI Network}

\Author{Xiaopeng Cheng $^{1,}$*$^{,\dagger}$\orcidA{}, Ilje Cho $^{2,}$*$^{,\dagger}$\orcidC{}, Tomohisa Kawashima $^{3,\dagger}$\orcidD{}, Motoki Kino $^{4,5,\dagger}$\orcidE{}, Guang-Yao Zhao $^{2}$\orcidF{}, Juan-Carlos Algaba $^{6}$\orcidG{}, Yutaro Kofuji $^{7,8}$, Sang-Sung Lee $^{1,9}$\orcidI{}, Jee-Won Lee$^{1}$\orcidJ{}, Whee Yeon Cheong $^{1,9}$\orcidK{}, Wu~Jiang~$^{10}$\orcidL{} and Junghwan Oh $^{11}$\orcidB{}
}

\AuthorNames{Xiaopeng Cheng, Ilje Cho, Tomohisa Kawashima, Motoki Kino, Guang-Yao Zhao, Juan-Carlos Algaba, Yutaro Kofuji, Sang-Sung Lee, Jee Won Lee, Whee Yeon Cheong, Wu Jiang  and Junghwan Oh}

\AuthorCitation{Cheng, X.; Cho, I.; Kawashima, T.; Kino, M.; Zhao, Gu.; Algaba, Ju.; Kofuji, Y.; Lee, Sa.; Lee, J.W.; Cheong, W.Y.; et al.}

\address{%
$^{1}$ \quad Korea Astronomy and Space Science Institute, Daedeok-daero 776, Yuseong-gu, Daejeon~34055,~Republic~of~Korea; fusion0417@gmail.com (S.-S.L.); jwlee78@kasi.re.kr (J.W.L.); wheeyeon@gmail.com (W.Y.C.)\\
$^{2}$ \quad Instituto de Astrof\'{\i}sica de Andaluc\'{\i}a---CSIC, Glorieta de la Astronom\'{\i}a s/n, E-18008~Granada, Spain; gyzhao@iaa.es\\
$^{3}$ \quad Institute for Cosmic Ray Research, The University of Tokyo, 5-1-5 Kashiwanoha, Kashiwa~277-8582, Chiba ,~Japan; kawshm@icrr.u-tokyo.ac.jp\\
$^{4}$ \quad National Astronomical Observatory of Japan, 2-21-1 Osawa, Mitaka, Tokyo~181-8588, Japan; motoki.kino@nao.ac.jp\\
$^{5}$ \quad Kogakuin University of Technology \& Engineering, Academic Support Center, 2665-1 Nakano, Hachioji, Tokyo 192-0015, Japan\\
$^{6}$ \quad Department of Physics, Faculty of Science, University of Malaya, Kuala Lumpur 50603, Malaysia; algaba@um.edu.my\\
$^{7}$ \quad Mizusawa VLBI Observatory, National Astronomical Observatory of Japan, 2-12 Hoshigaoka, Mizusawa, Oshu, Iwate 023-0861, Japan; kofuji-yutaro011@g.ecc.u-tokyo.ac.jp\\
$^{8}$ \quad Mizusawa VLBI Observatory, National Astronomical Observatory of Japan, 2-12 Hoshigaoka, Mizusawa, Oshu, Iwate 023-0861, Japan\\
$^{9}$ \quad University of Science and Technology, Gajeong-ro 217, Yuseong-gu, Daejeon 34113, Republic of Korea\\
$^{10}$ \quad Shanghai Astronomical Observatory, Chinese Academy of Sciences, Nandan Road 80, Shanghai~200030,~China; jiangwu@shao.ac.cn \\
$^{11}$ \quad Joint Institute for VLBI ERIC, Oude Hoogeveensedijk 4, 7991 PD Dwingeloo, The Netherlands; oh@jive.eu \\
}
\corres{Correspondence: xcheng@kasi.re.kr (X.C.); icho@iaa.es (I.C.)}

\firstnote{These authors contributed equally to this work.}

\abstract{
In this work, we studied the Galactic Center supermassive black hole (SMBH), Sagittarius A* (\sgra), with the KVN and VERA Array (KaVA)/East Asian VLBI Network (EAVN) monitoring observations. 
Especially in 13 May 2019, \sgra experienced an unprecedented bright near infra-red (NIR) flare; so, we find a possible counterpart at 43\,GHz (7\,mm). As a result, a large temporal variation of the flux density at the level $\sim$15.4\%, with the highest flux density of 2.04 Jy, is found on 11 May 2019. Interestingly, the intrinsic sizes are also variable, and the area and major-axis size show marginal correlation with flux density with $\gtrsim$2\,$\sigma$. Thus, we {interpret} that the emission region at 43~GHz follows the larger-when-brighter relation in 2019. The possible origins are discussed with an emergence of a weak jet/outflow component and the position angle change of the rotation axis of the accretion disk in time.
}


\begin{document}


\section{Introduction}

\textls[-15]{The supermassive black hole (SMBH) in our Galactic Center, Sagittarius A* (\sgra), is the closest known SMBH with a mass $M_{\rm BH}$$\sim$$4\times10^6 M_{\odot}$ (e.g., \citep{genzel2010,gravity2020}) at a distance D$\sim$8.1~kpc~\citep{gravity2019}}.
Thanks to its proximity, \sgra subtends the largest angular size (Schwarzschild radius, $\rm R_{s}$$\sim$10 $\upmu$as) on the sky among all known black holes and is one of the most promising targets for the Event Horizon Telescope (EHT) to study the vicinity of a black hole through direct imaging. 

Recently, the~EHT Collaboration published the \sgra black hole shadow results at $\lambda$ = 1.3 mm, showing angular diameter d$\rm _{sh}$ = 48.7 $\pm$ 7.0 $\upmu$as with a bright and thick emission ring of a diameter $\theta$ = 51.8 $\pm$ 2.3 $\upmu$as~\citep{2022ApJ...930L..12E,2022ApJ...930L..13E,2022ApJ...930L..14E,2022ApJ...930L..15E,2022ApJ...930L..16E,2022ApJ...930L..17E}. 
However, the~emission mechanism of \sgra is being debated as to whether the jet base (e.g.,~\citep{2000A&A...362..113F}) or a radiatively inefficient accretion flow (RIAF)~(e.g.,~\citep{2003ApJ...598..301Y}).

Unlike the extragalactic active galactic nuclei (AGNs), the~relativistic jet feature has not been observed yet in \sgra with the Very Long Baseline Interferometry (VLBI) observations. 
Especially at centimeter (cm) wavelengths, the~source structure is dominated by scatter broadening caused by the ionized interstellar scattering medium (ISM) along the line of sight~\citep{1992RSPTA.341..151N,2006ApJ...648L.127B}; thus, it is difficult to resolve the fine structure. 
However, as~the observed size follows the relation of square of observing wavelength, $\lambda$, the~effect becomes weaker at shorter wavelengths so that the intrinsic source structure can be visible~\citep{2004Sci...304..704B}.

\sgra also shows variability from minutes to months timescale at a variety of wavelengths~(e.g.,~\citep{2003Natur.425..934G,2013ApJ...774...42N}). 
The rapid increase in near-infrared (NIR) flux is seen several times per day~\citep{2003Natur.425..934G}.
The brighter NIR flares are often associated with the X-ray flares after a few tens of minutes, but~there are numerous NIR flares without an X-ray counterpart~\citep{2006A&A...450..535E,2012ApJ...758L..11Y,2017MNRAS.468.2447P,2018ApJ...864...58F}.
The millimetre (mm)/sub-mm emission is much more stable~\citep{2008ApJ...682..373M}.
Recently, \citet{2019ApJ...882L..27D} reported that the NIR peak flux levels were brighter in 2019 April 20 than 99.7\% of all historical data points, and~an NIR flare of unprecedented brightness in 2019 May 13 with flux peaks ($\sim$6 mJy) that are twice the values from previous measurements. 
The flare has shown rapid, continuous decrease from the peak to $\sim$1 mJy in 1 h.
They also find the flux variations observed in 2019 to be significantly different than in the historical data from \citet{2018ApJ...863...15W}. 
They suggest that this may indicate that \sgra is experiencing a physical change in the accretion activity, possibly due to the pericenter passage of the star S0-2 in 2018 or the gaseous object G2 in 2014~\citep{2013A&A...551A..18E,2019Sci...365..664D}.
However, \citet{2018MNRAS.478.3544R} have argued that the effect of S0-2 on the RIAF structure should be negligible.
As for the G2, on~the other hand, a longer timescale (5--10 years) between the increased mass accretion and magnetic energy/flux enhancement has been predicted~\citep{2017PASJ...69...43K}, so its impact may still be valid to investigate.

To look for the possible imprints of G2 encounter, the~Korean VLBI Network (KVN\endnote{Korean VLBI Network, which consists of three 21\,m telescopes in Korea: Yonsei (KYS), Ulsan (KUS), and~Tamna (KTN).}) and VLBI Exploration of Radio Astrometry (VERA\endnote{VLBI Exploration of Radio Astrometry, which consists of four 20\,m telescopes in Japan: Mizusawa (MIZ), Iriki (IRK), Ogasawara (OGA), and~Ishigakijima (ISG).}) Array, KaVA, started regular observations of \sgra, as~one of the main targets of the large AGN program, at~22 and 43 GHz from March 2014~\citep{2014IAUS..303..288A,2014AJ....147...77L,kino15}. 
{In the second half of 2018, the~East Asian VLBI Network (EAVN) also started its open-use program} ~\citep{2016ASPC..502...81W,2018NatAs...2..118A,2021RAA....21..205C}.
The EAVN campaign was performed by making use of the slots allocated to the KaVA AGN Large Program that intensively monitored the nearby SMBHs, M\,87 and \sgra, at~22 and 43\,GHz. 
From our long-term monitoring results since 2014, the~highest flux density of \sgra at 43\,GHz ($\sim$2.04 Jy) was detected in 2019 May, which may be related to the NIR flare and possibly the G2 encounter.

In this work, therefore, we report the accurate measurements of flux density and intrinsic size of \sgra from the KaVA/EAVN observations at 43\,GHz in 2019.
In Section~\ref{2}, we present the observations and data analysis. Section~\ref{3} presents the light curves, size measurement, and~correlation test. Section~\ref{4} presents the discussion of potential physical explanations for these~observations.

\section{Observations and Data~Analysis}
\label{2}
\unskip

\subsection{Observations}

\begin{figure}[H]
    \includegraphics[height=8cm]{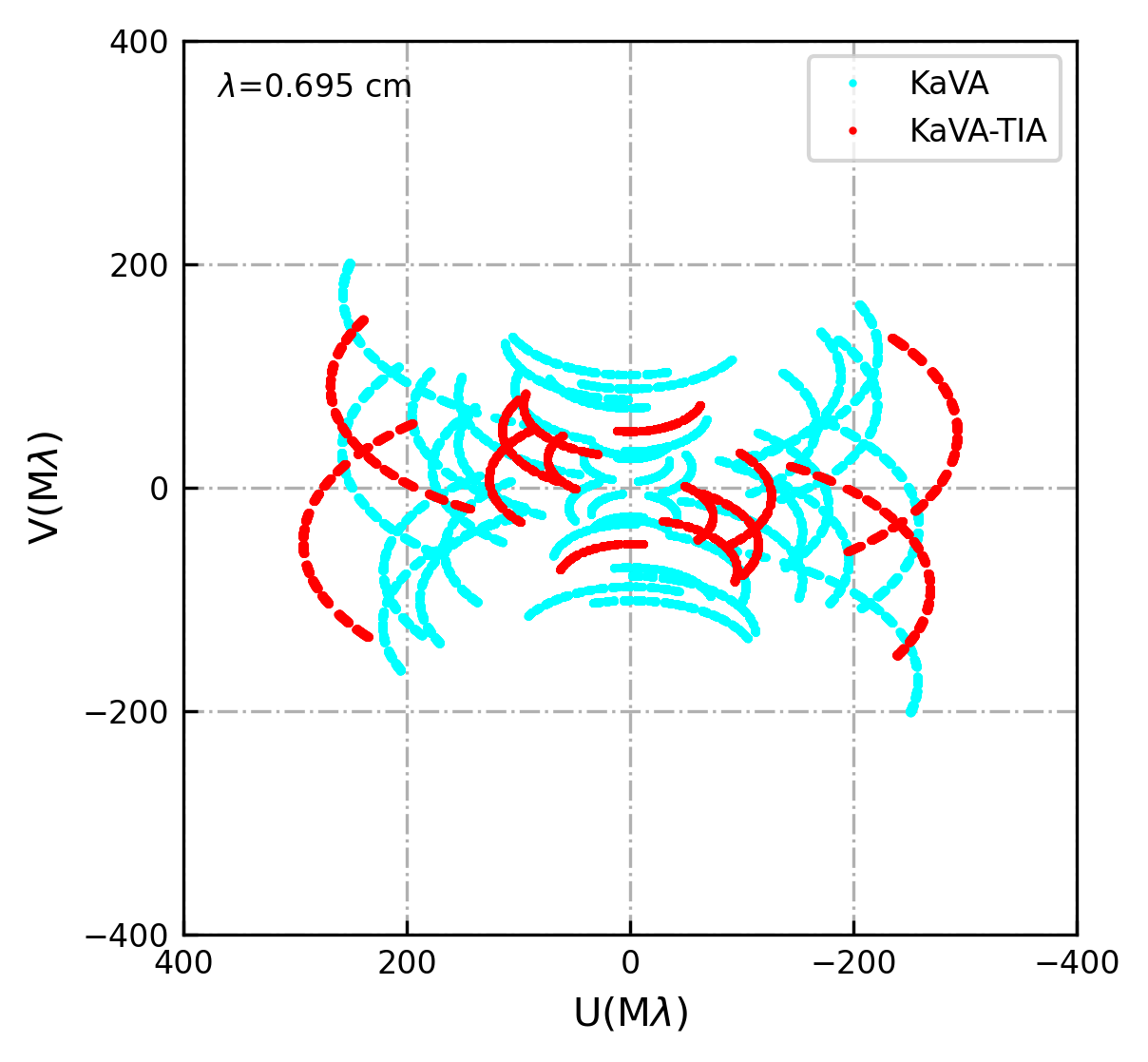}
    \caption{Typical $u-v$ coverage of EAVN observations (a19kh01c) at 43 GHz. Each point has been averaged within 30 s. TIA provides the longest baselines (red points) in an east-west~direction.} 
    \label{fig:uvplot}
\end{figure}

Twelve observations were performed at 43 GHz with the EAVN as part of the KaVA/ EAVN AGN large program~\citep{kino15} in 2019.
The participating stations are KaVA and 2 additional East Asian telescopes (Tianma 65m and Nobeyama 45m;~\citep{2021RAA....21..205C}).
Three of the observations failed due to bad weather or station maintenance.
This study is mainly based on the nine successful observations, as listed in Table~\ref{log}. 
The data were recorded with 256 MHz total bandwidth in left-handed circular polarization (LCP), resulting in a sampling rate of 1 Gbps.
{While} three observations (a19kh01a, a19kh01c, and~a19kh01f) {were recorded with} 32~MHz~$\times$~8 intermediate frequencies (IFs) {band}, the other six observations used 16~MHz~$\times$~16~IFs.
Each observation lasted for about 6 h and the on-source time for \sgra and the main calibrator NRAO 530 was about 200 and 30 min, {respectively}.
Figure~\ref{fig:uvplot} displays {an example of EAVN} $(u, v)$ coverage towards \sgra at 43\,GHz. 
The correlation was carried out in the Korea–Japan Correlation Center (KJCC) at Daejeon, Korea~\citep{2014AJ....147...77L}.
Tianma 65m (TIA) participated in seven (out of nine) observations, and~two of them (a19mk01a and a19mk01c) had no fringes due to frequency setup.
Nobeyama 45m (NRO45) participated in one observation (a19kh01c), but~no fringes were detected.

\begin{table}[H]
\small
\caption{Summary of the EAVN~observations\label{log}}
\setlength{\tabcolsep}{0.15cm}
\begin{adjustwidth}{-\extralength}{0cm}
\newcolumntype{C}{>{\centering\arraybackslash}X}
		\begin{tabularx}{\fulllength}{CCcCCC}
\toprule
\textbf{Project Code}& \textbf{Experiment Date} &  \textbf{Participating Stations} \boldmath{$^{a}$} & \textbf{Image rms} \boldmath{$^{b}$}   \textbf{mJy~beam}\boldmath{$^{-1}$}    & \textbf{Peak Intensity} \boldmath{$^{c}$} \textbf{Jy~beam}\boldmath{$^{-1}$} & \boldmath{$\theta_{\mathrm{FWHM}}$} \textbf{(mas,~mas,~deg)} \\
\textbf{ (1)}  & \textbf{(2)}        & \textbf{(3) }                    & \textbf{(4)}             & \textbf{(5)}            &  \textbf{(6)}         \\
\midrule
a19mk01a & 2019-02-27 & KaVA ($-$IRK) + TIA& 2.01  & 1.240 & (1.33, 0.79, $-$5.7)  \\
a19kh01a & 2019-03-22 & KaVA + TIA         & 1.78  & 0.624 & (1.37, 0.37, $-$16.3) \\
a19kh01c & 2019-03-29 & KaVA + TIA + NRO45 & 1.41  & 0.579 & (1.25, 0.38, $-$17.8) \\
a19kh01f & 2019-04-12 & KaVA + TIA ($-$ KYS)& 1.69 & 0.712 & (1.01, 0.37, $-$16.6) \\
a19mk01c & 2019-05-11 & KaVA + TIA          & 1.96 & 0.961 & (1.24, 0.43, $-$16.3) \\
a19mk01e & 2019-09-10 & KaVA ($-$ KYS)      & 2.56 & 1.039 & (1.37, 0.62, $-$8.3) \\
a19mk01g & 2019-10-10 & KaVA + TIA ($-$ OGA) & 1.89 & 0.564 & (1.77, 0.42, $ -$17.3) \\
a19mk01h & 2019-11-23 & KaVA               & 2.10 & 0.764 & (1.32, 0.37, $-$9.9)  \\
a19mk01i & 2019-12-18 & KaVA + TIA         & 1.55 & 0.861 & (1.48, 0.65, $-$6.5)  \\
\bottomrule
\end{tabularx}
\end{adjustwidth}
Note. $^{a}$ In brackets, ($-$) means the KaVA stations that were not used in individual observation, and~($+$) means the non-KaVA but EAVN stations that participated in the observation. 
$^{b}$ The image sensitivity is based on the model fitting and self-calibration.
$^{c}$ The peak intensity is based on the uniform weighting and the corresponding synthesized beam.
\end{table}
\unskip

\subsection{Data~Reduction}
\label{2.2}

The data were calibrated with the NRAO Astronomical Image Processing System (AIPS) software package~\citep{2003ASSL..285..109G}.
Firstly, the~sampler voltage offsets were corrected by auto-correlation and a multiplicative correction factor of 1.3 was applied to all data to correct the quantization loss in the Daejeon hardware correlator~\citep{2015JKAS...48..229L}.
The a-priori amplitude calibrations (APCAL) were conducted using the information of the gain curve (GC) and system temperature (TY) for KaVA and NRO45 telescopes.
Because of the large amplitude offset for TIA, we used the template spectrum method (ACFIT) for the amplitude calibration by using the SiO maser lines from OH\,0.55--0.06 and VX\,Sgr, which can give a more realistic antenna gain curve as a function of the elevation than the a priori calibration method~\citep{2017PASJ...69...87C}.
The phase contributions from the antenna parallactic angles were removed before any other phase corrections were applied.
The station KUS was chosen as the reference antenna. 
After removing instrumental phase offsets from each IF by using NRAO 530 (i.e., manual PCAL), the~fringe-fitting and bandpass calibration were conducted directly to the \sgra visibilities. 
The visibilities at lower elevations $<$5$\degree$ at either telescope of a baseline were flagged.
We also excluded the first and sixteenth IFs for 16 MHz $\times$ 16 IFs mode data because of their very low correlation amplitude. 
Finally, the~data were averaged over all IFs and 30 s and split into single-source~files.

Before model fitting, we first add the fractional systematic error to the data, which inflates the thermal noises.
This is mainly to account for the non-closing error budget, as~well as to avoid biases in the fitted model by the uncalibrated station gains (e.g., 10\%). 
For instance,  the determined overall telescope gain correction factors were found to be small in NRAO 530, typically within 10\%, in~agreement with the typical mm VLBI observations (Table~\ref{gain}). For TIA data, on the other hand, we add 30\% of the visibility amplitude as the systematic error to account for additional uncertainties (e.g., from~residual bandpass, system temperature measurements, and~pointing;~\citep{2021RAA....21..205C}).

\begin{table}[H]
\caption{\xc{Amplitude gain correction factors for each~station\label{gain}}}
\setlength{\tabcolsep}{0.15cm}
\newcolumntype{C}{>{\centering\arraybackslash}X}
\begin{tabularx}{\textwidth}{ccCCCCCCCC}
\toprule
\textbf{Epoch} &\textbf{ Source} &  \textbf{KTN} & \textbf{KYS} &\textbf{ KUS} & \textbf{OGA} & \textbf{MIZ} & \textbf{ISG} & \textbf{IRK} & \textbf{TIA} \\
\textbf{(1)}   & \textbf{(2)}    & \textbf{(3)}  & \textbf{(4)} & \textbf{(5)} & \textbf{(6)} & \textbf{(7)} & \textbf{(8)} & \textbf{(9)} & \textbf{(10)}  \\
\midrule
a19mk01a & \sgra    & 0.96 & 1.01 & 1.01 & 0.86 & 1.08 & 1.00 &  &  \\
         & NRAO 530 & 0.99 & 1.01 & 0.97 & 1.00 & 1.02 & 1.00 &  &  \\
a19kh01a & \sgra    & 0.96 & 1.03 & 0.94 & 0.91 & 1.07 & 1.01 & 1.03 & 1.19 \\
         & NRAO 530 & 0.94 & 1.03 & 0.94 & 0.97 & 0.99 & 1.01 & 0.99 & 1.25 \\
a19kh01c & \sgra    & 0.89 & 1.17 & 1.00 & 1.13 & 1.06 & 0.98 & 0.90 & 1.32 \\
         & NRAO 530 & 0.89 & 1.04 & 0.94 & 1.01 & 0.99 & 1.01 & 0.93 & 1.36 \\
a19kh01f & \sgra    & 0.96 &      & 0.98 & 0.92 & 1.14 & 1.01 & 0.96 &  \\
         & NRAO 530 & 0.94 &      & 0.97 & 1.02 & 1.01 & 1.02 & 0.98 &  \\
a19mk01c & \sgra    & 1.01 & 1.02 & 1.00 & 1.07 & 0.99 & 0.97 & 0.98 &  \\
         & NRAO 530 & 0.99 & 1.01 & 0.97 & 1.01 & 1.01 & 1.02 & 0.98 &  \\
a19mk01e & \sgra    & 1.00 &      & 0.98 & 1.00 & 1.02 & 0.99 & 1.01 &  \\
         & NRAO 530 & 0.94 &      & 0.98 & 1.00 & 1.02 & 1.03 & 0.96 &  \\
a19mk01g & \sgra    & 1.01 & 0.89 & 0.97 &      & 1.23 & 1.05 & 0.89 & 1.11 \\
         & NRAO 530 & 0.98 & 0.96 & 0.98 &      & 1.02 & 1.03 & 1.01 & 1.23 \\
a19mk01h & \sgra    & 1.01 & 0.97 & 1.01 & 0.98 & 1.10 & 0.94 & 1.01 &  \\
         & NRAO 530 & 1.03 & 0.91 & 0.97 & 1.03 & 1.02 & 1.04 & 1.04 &  \\
a19mk01i & \sgra    & 1.00 & 0.99 & 1.01 & 0.98 & 1.03 & 1.01 & 1.00 &  \\
         & NRAO 530 & 1.06 & 0.98 & 0.90 & 0.88 & 0.95 & 0.88 & 0.96 &  \\
\bottomrule
\end{tabularx}
\end{table}
\unskip

\subsection{Self-Calibration with a Gaussian~Model}
\label{2.3}

\begin{figure}[H]
    \includegraphics[height=2.8cm]{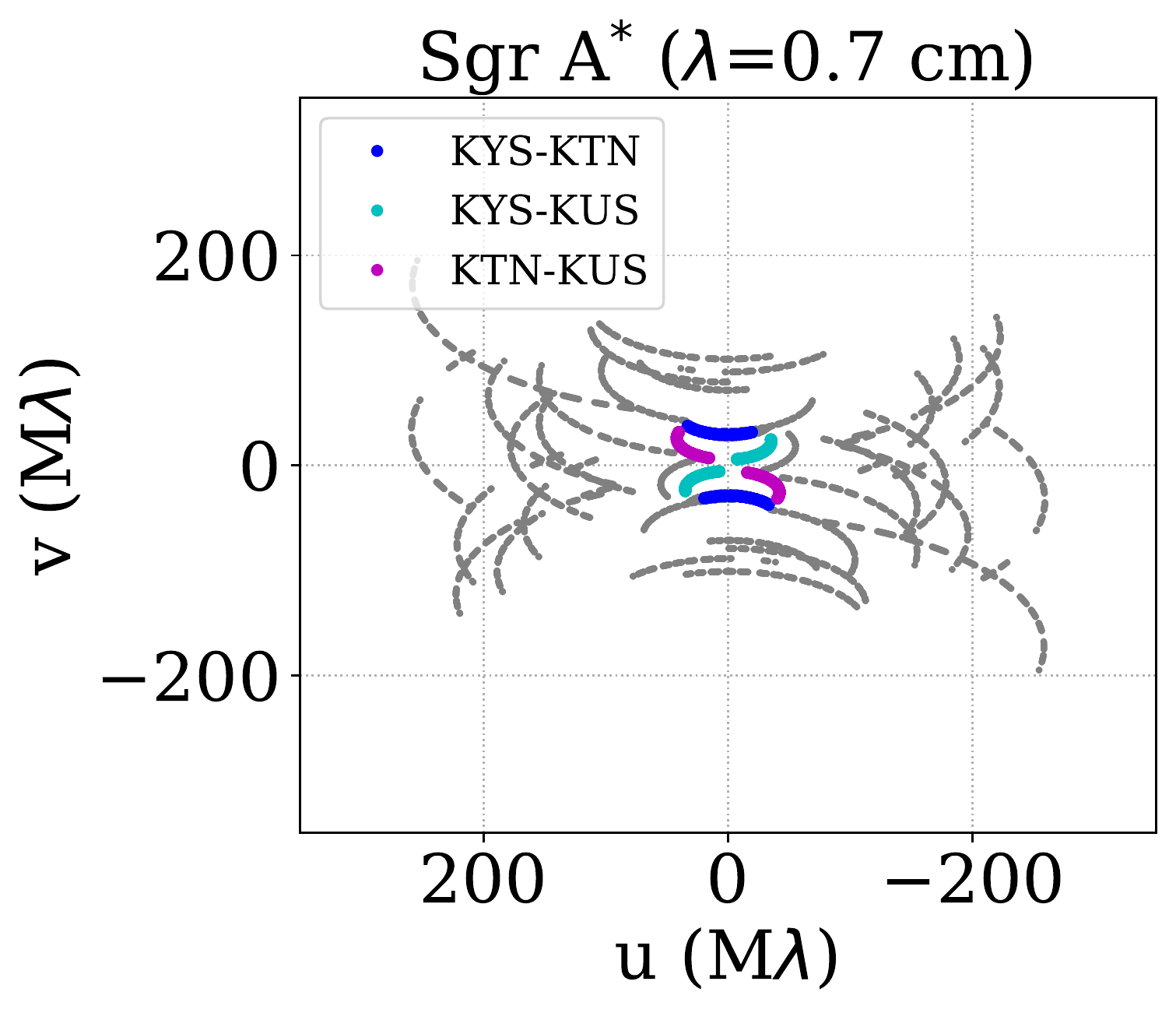}
    \includegraphics[height=2.8cm]{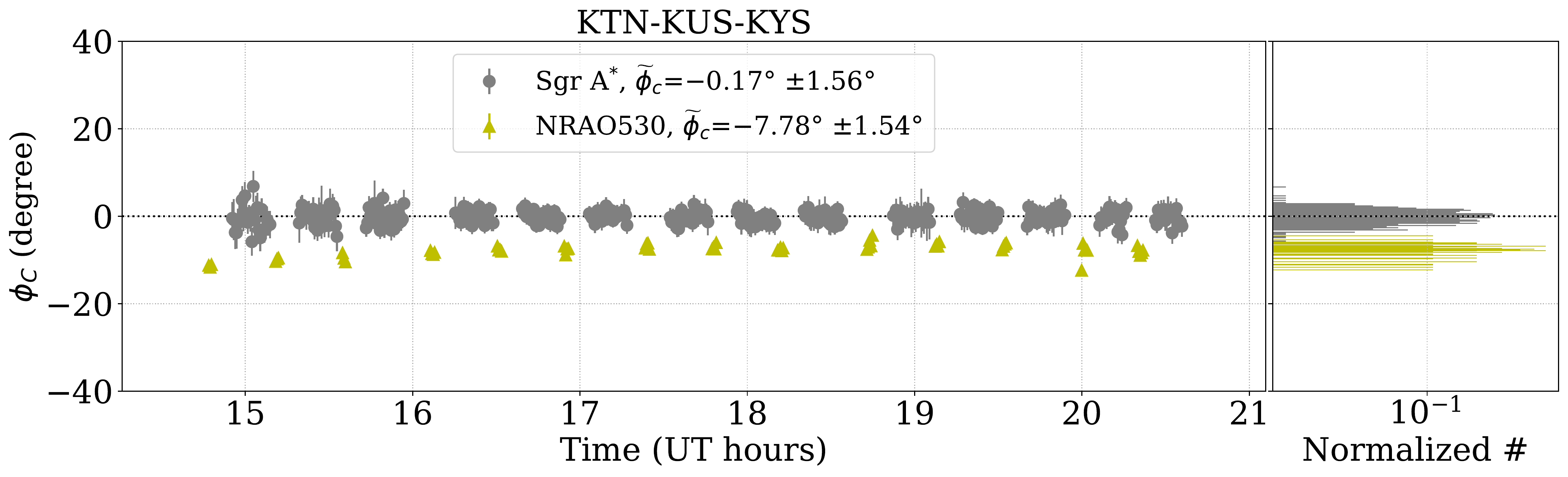}
    \\
    \includegraphics[height=2.8cm]{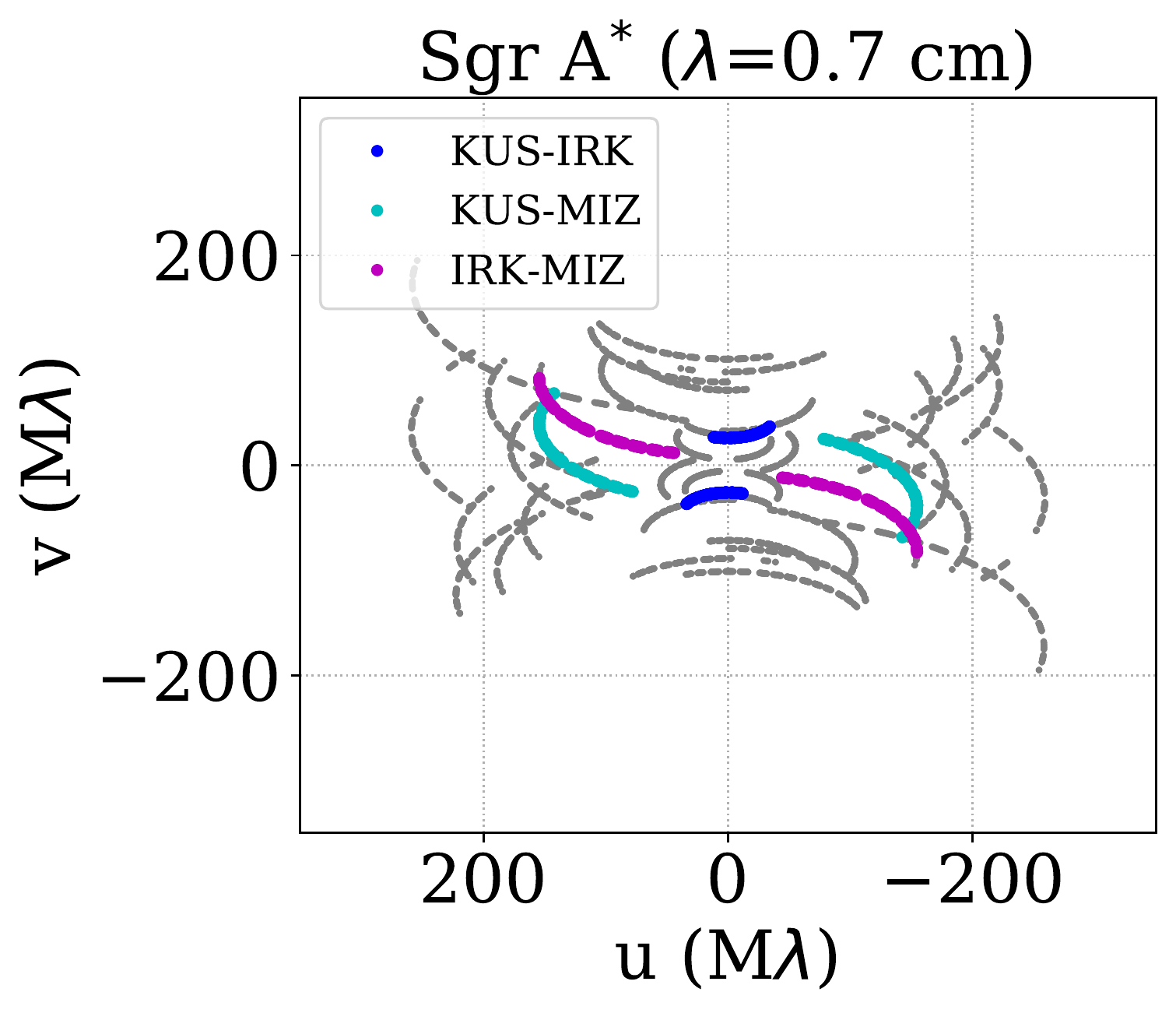}
    \includegraphics[height=2.8cm]{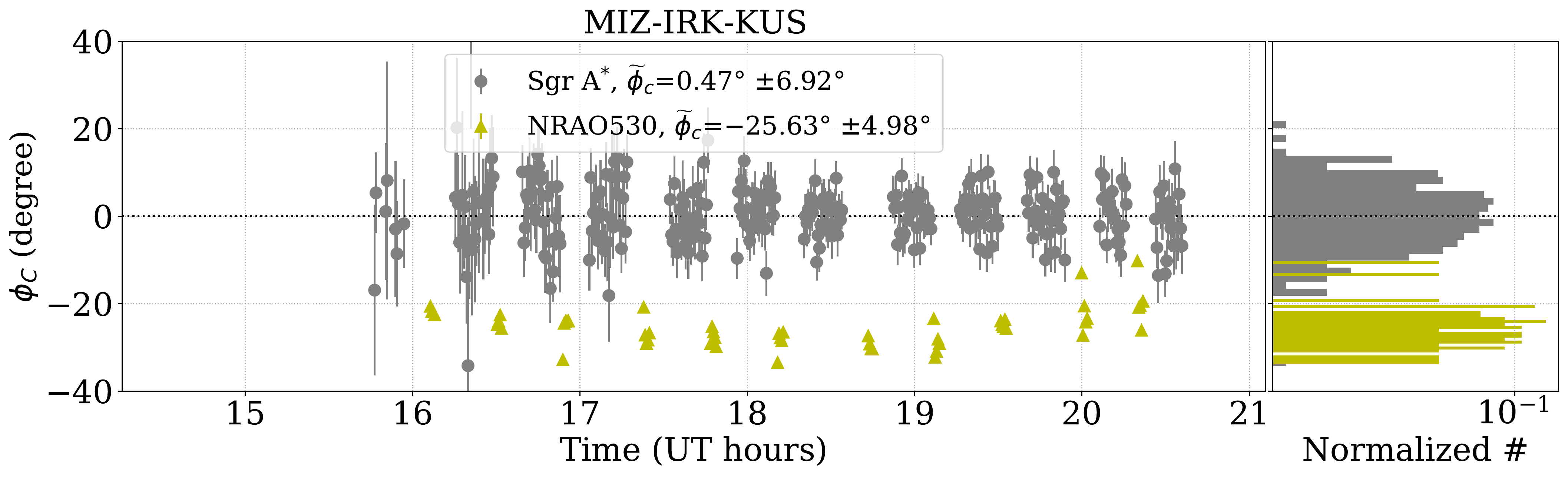}
    \\
    \includegraphics[height=2.8cm]{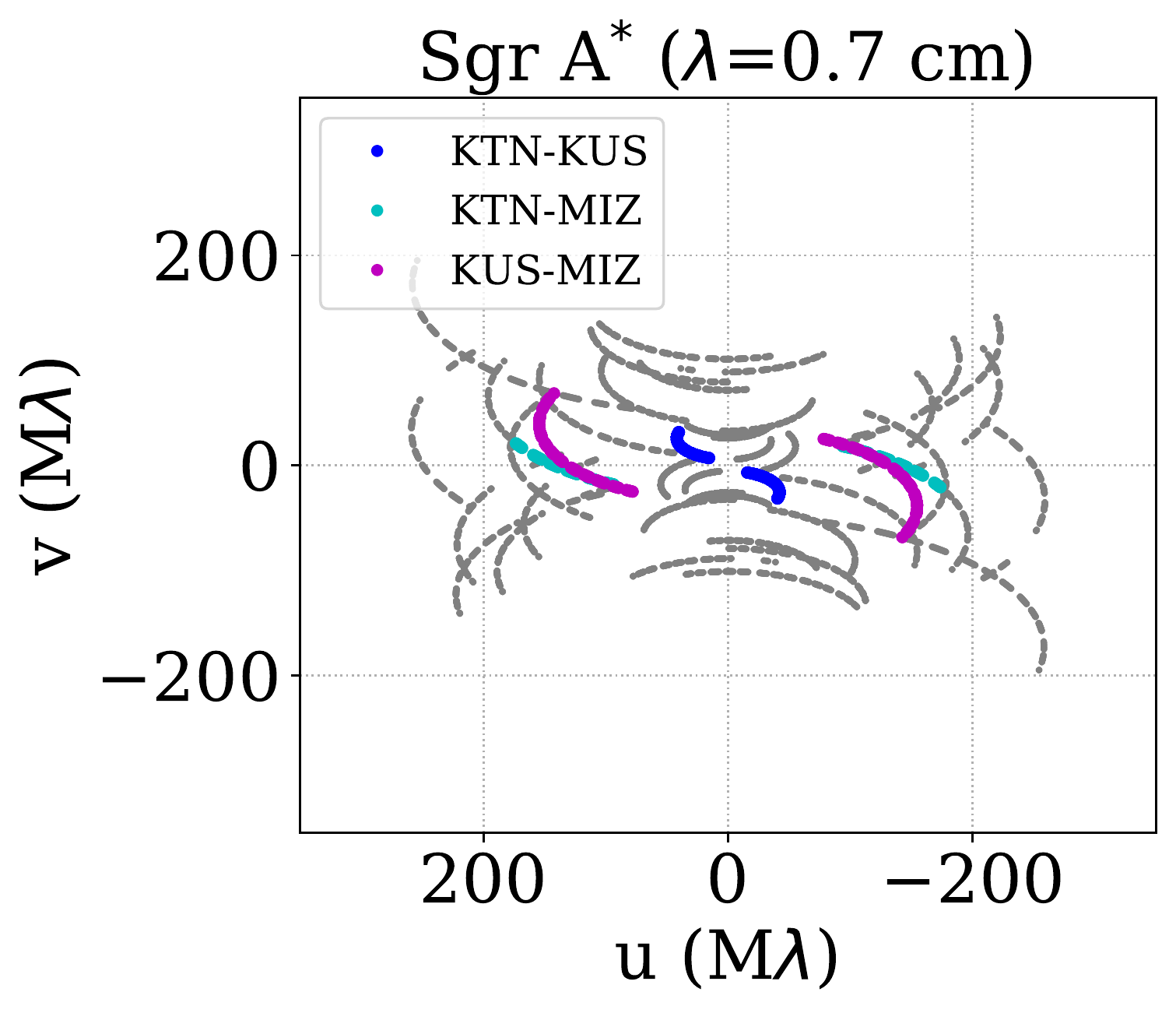}
    \includegraphics[height=2.8cm]{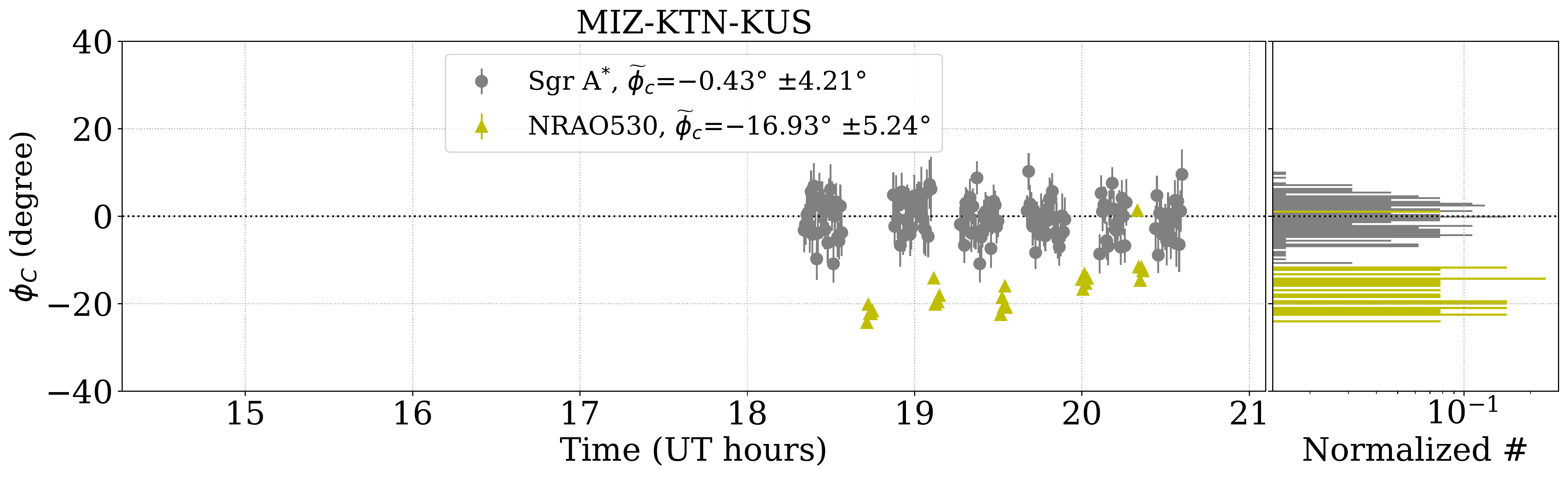}
    \\
    \includegraphics[height=2.8cm]{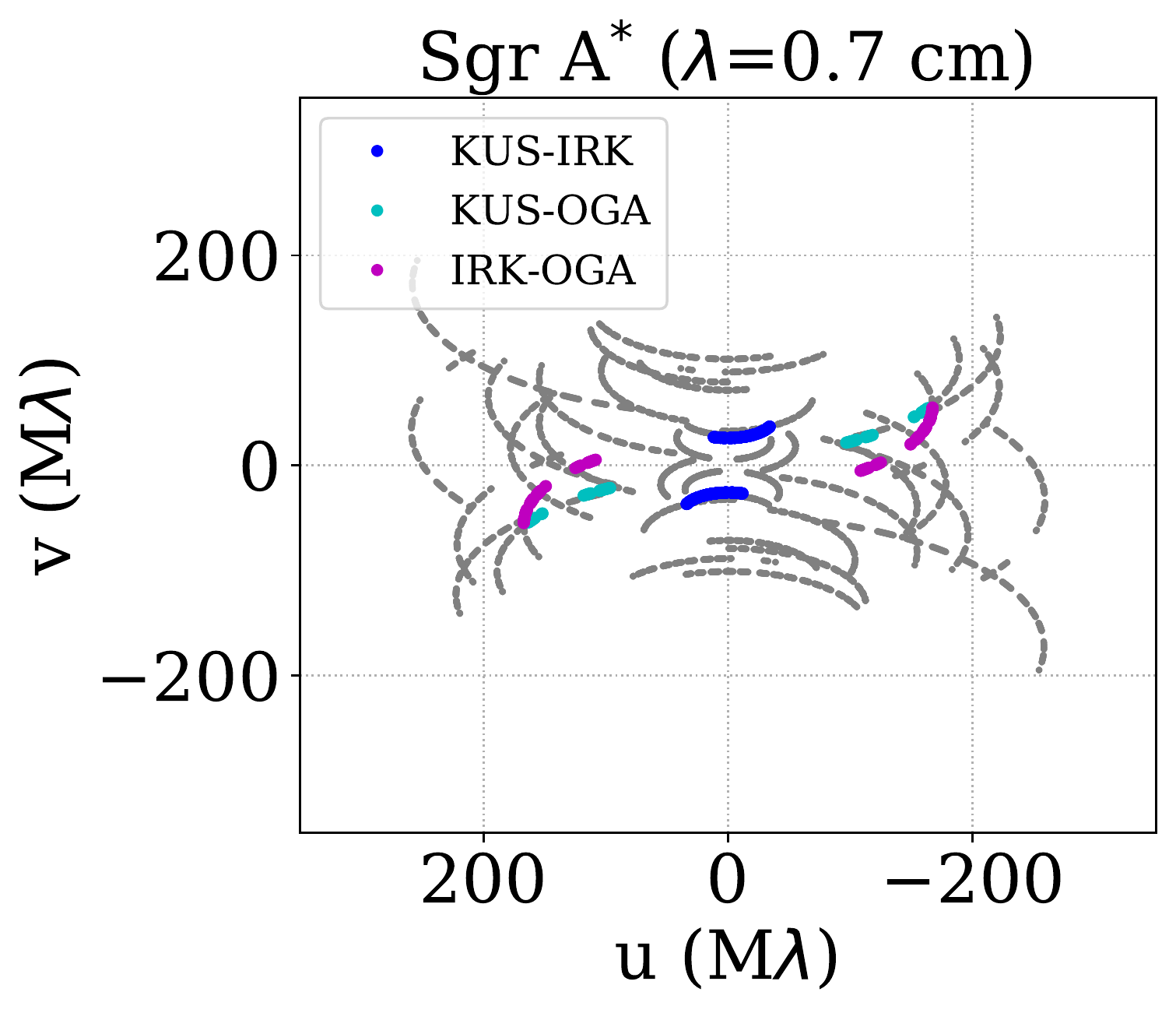}
    \includegraphics[height=2.8cm]{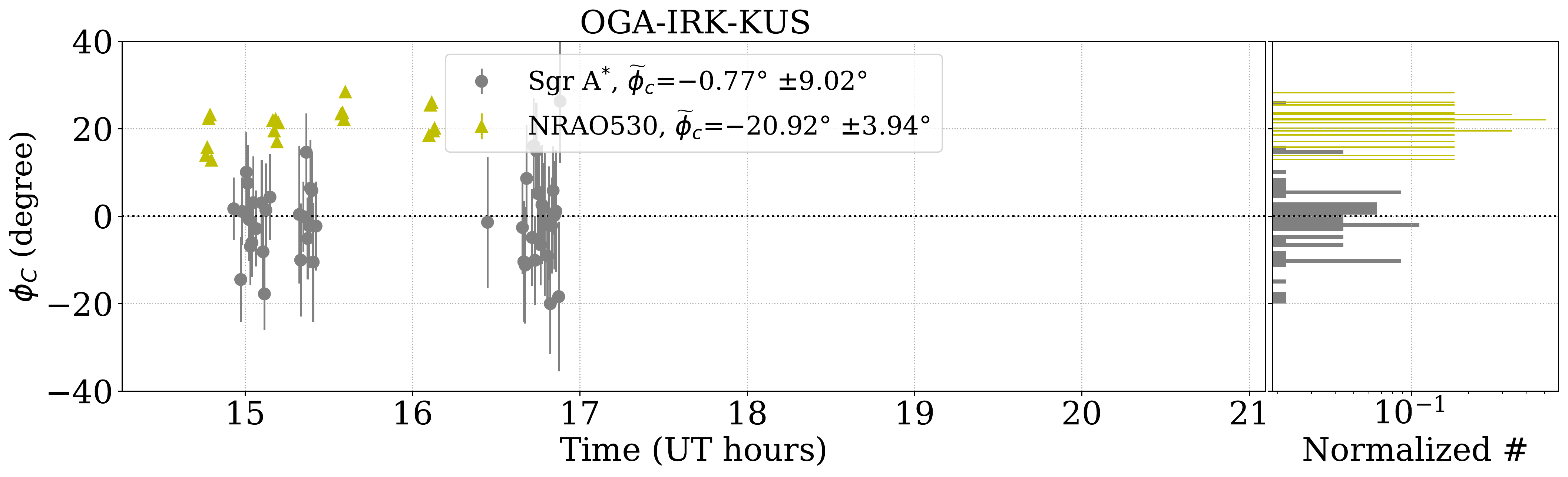}
    \\
    \includegraphics[height=2.8cm]{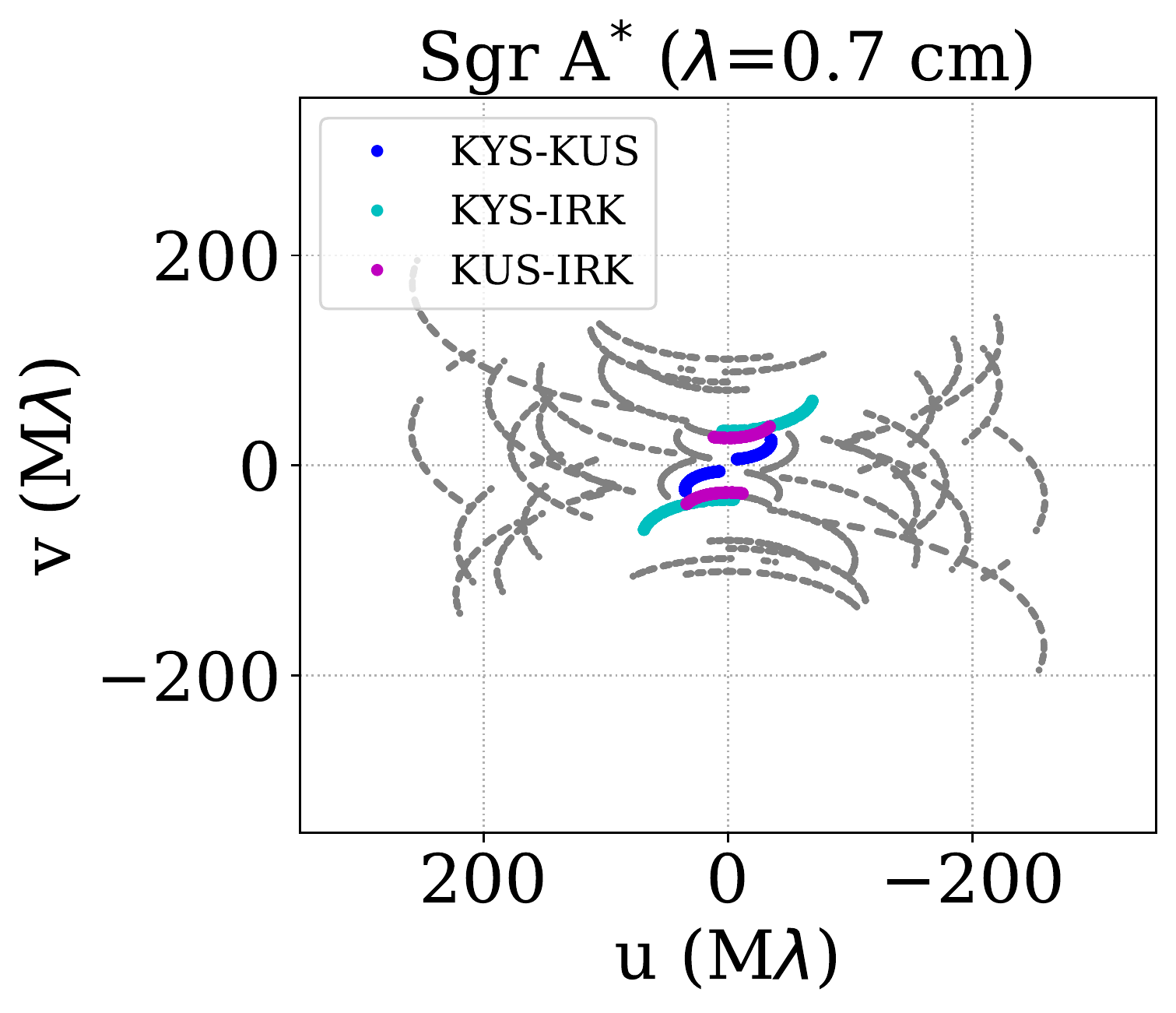}
    \includegraphics[height=2.8cm]{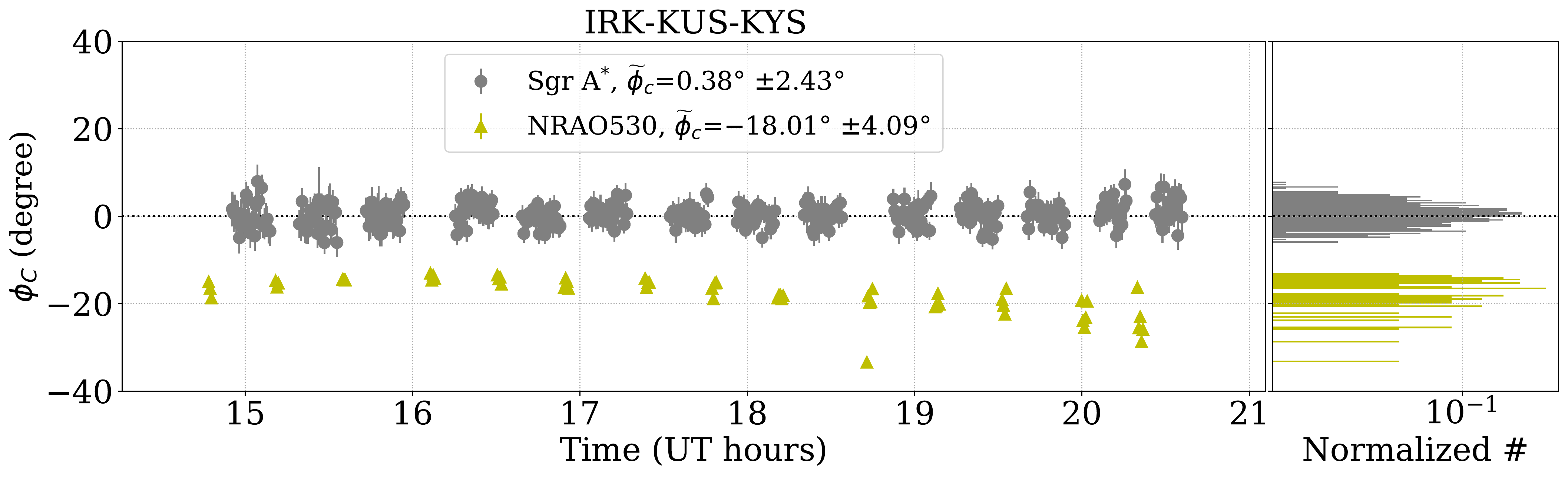}
    \\
    \caption{Closure phases of \sgra (gray) and NRAO~530 (yellow). From~left to right, the~corresponding baselines toward \sgra in the u$-$v plane (colored points), closure phases as a function of observing time, and~a histogram of the closure phases are shown. 
    The 30\,s averaged are shown, and~the error bar represents the quadratic sum of the thermal noises (i.e., without~the fractional systematic error) over the visibility amplitudes. 
    The time-averaged closure phase, $\widetilde\phi_{c}$, is shown in each legend, together with the uncertainty from the standard deviation. 
    }
    \label{CP}
\end{figure}

First, we checked the closure phases of \sgra and confirmed that 
{they are mostly distributed around zero degrees (i.e., no clear evidence of non-zero deviation; see Figure~\ref{CP}). This implies that \sgra can be reasonably modeled with a symmetric structure (e.g., a~single Gaussian). 
Based on this, we have fitted the data with an elliptical Gaussian model and self-calibrated the complex visibility with the result using DIFMAP~\citep{1997ASPC..125...77S}. 
Note that we have only used the visibilities where the baseline lengths are shorter than 176 M$\uplambda$ (i.e., ensemble-average image; e.g.,~\citep{2022ApJ...926..108C}). 
This is to avoid biases by the scattering effects, especially at long baselines where the refractive scattering noises get larger. 
}

The cm$-$mm VLBI observations toward \sgra is largely affected by the ISM as it looks through the Galactic Center. 
Therefore, the observed result itself is a combination of the intrinsic property of \sgra with the scattering effects, mainly the diffractive and refractive scattering. 
The diffractive scattering provides a Gaussian blurring, which is dominant at the ``short'' baseline lengths in the visibility domain. 
Due to the effect, the~observed size of \sgra is roughly the quadratic sum of its intrinsic size and the size of the scattering kernel (e.g.,~\citep{2011A&A...525A..76L}), and~is proportional to the square of the observing wavelength. 
On the other hand, the~refractive scattering introduces non-Gaussian sub-structures (in the image domain) and complex noises (in the visibility domain) which are dominant at long baseline lengths. 
Here, the baseline length is determined by $(1+D/R)r_{\rm in}$, where $D$, $R$, and~$r_{\rm in}$ are 
the distance between Earth and the scattering screen, the distance between \sgra and the scattering screen, and a finite inner scale of interstellar turbulence, respectively.  
With $D=2.7\,$kpc, $R=5.4\,$kpc, $r_{\rm in}=800\,$km~\citep{2018ApJ...865..104J}, therefore, the~``short'' baseline length corresponds to $176\,{\rm M}\lambda$ at 43\,GHz so we have applied this threshold to avoid the effects of refractive noise in our model fitting. 
In addition, we have flagged the data with a signal-to-noise ratio (S/N) lower than 3 (for thermal noise) and 4 (for refractive noise) (e.g.,~\citep{2018ApJ...865..104J}). 
Note that the refractive noises are derived following the previous study \citep{2022ApJ...926..108C}.

\begin{figure}[H]
    \includegraphics[height=8cm]{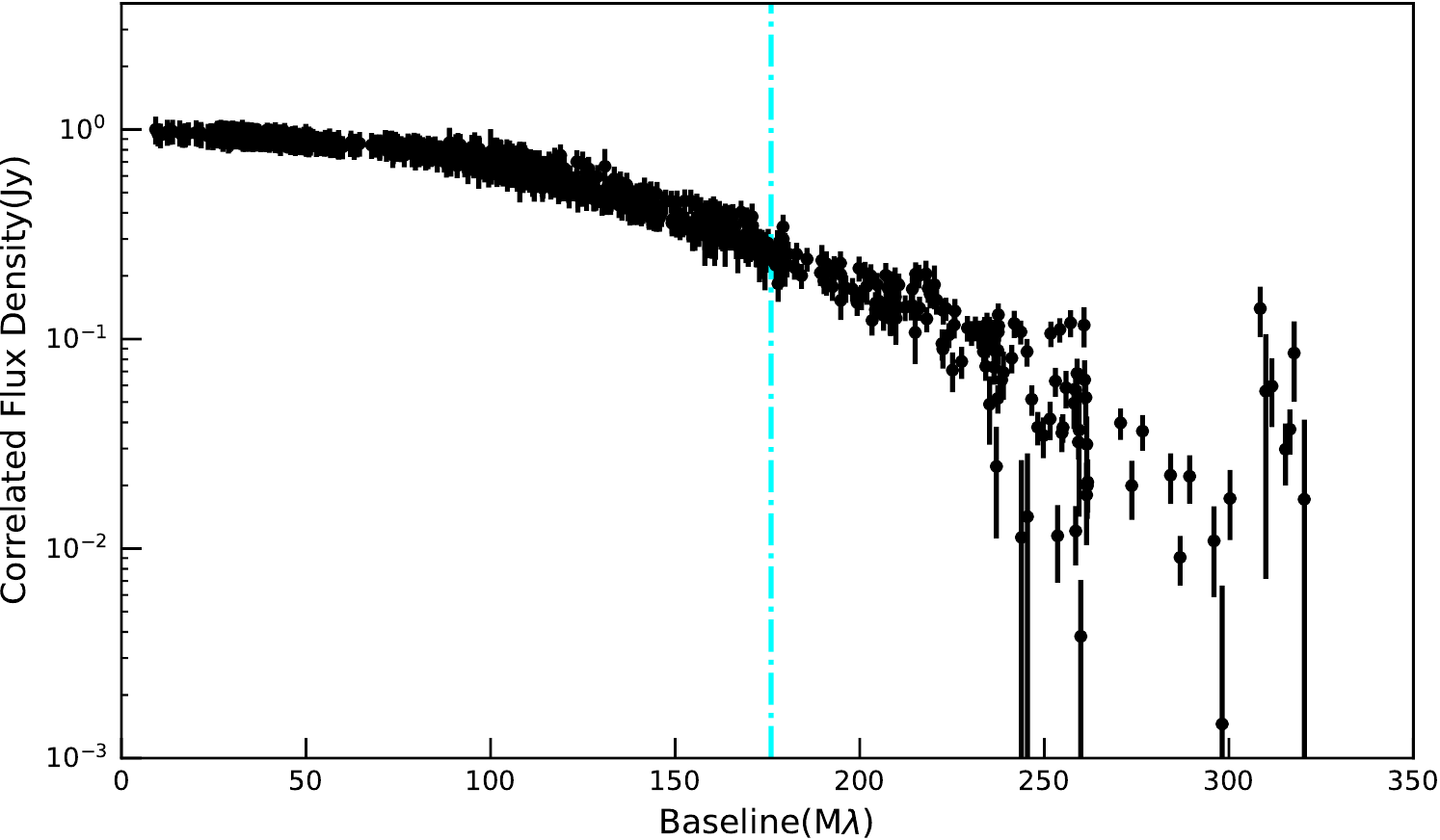} 
    \caption{Normalized self-calibrated correlated flux density of \sgra at 43 GHz. The~cyan-colored vertical line shows the  ``short'' range. Each point has been 5 min averaged.}
    \label{fig:radplot}
\end{figure}

\begin{figure}[H]
    \includegraphics[height=5.5cm]{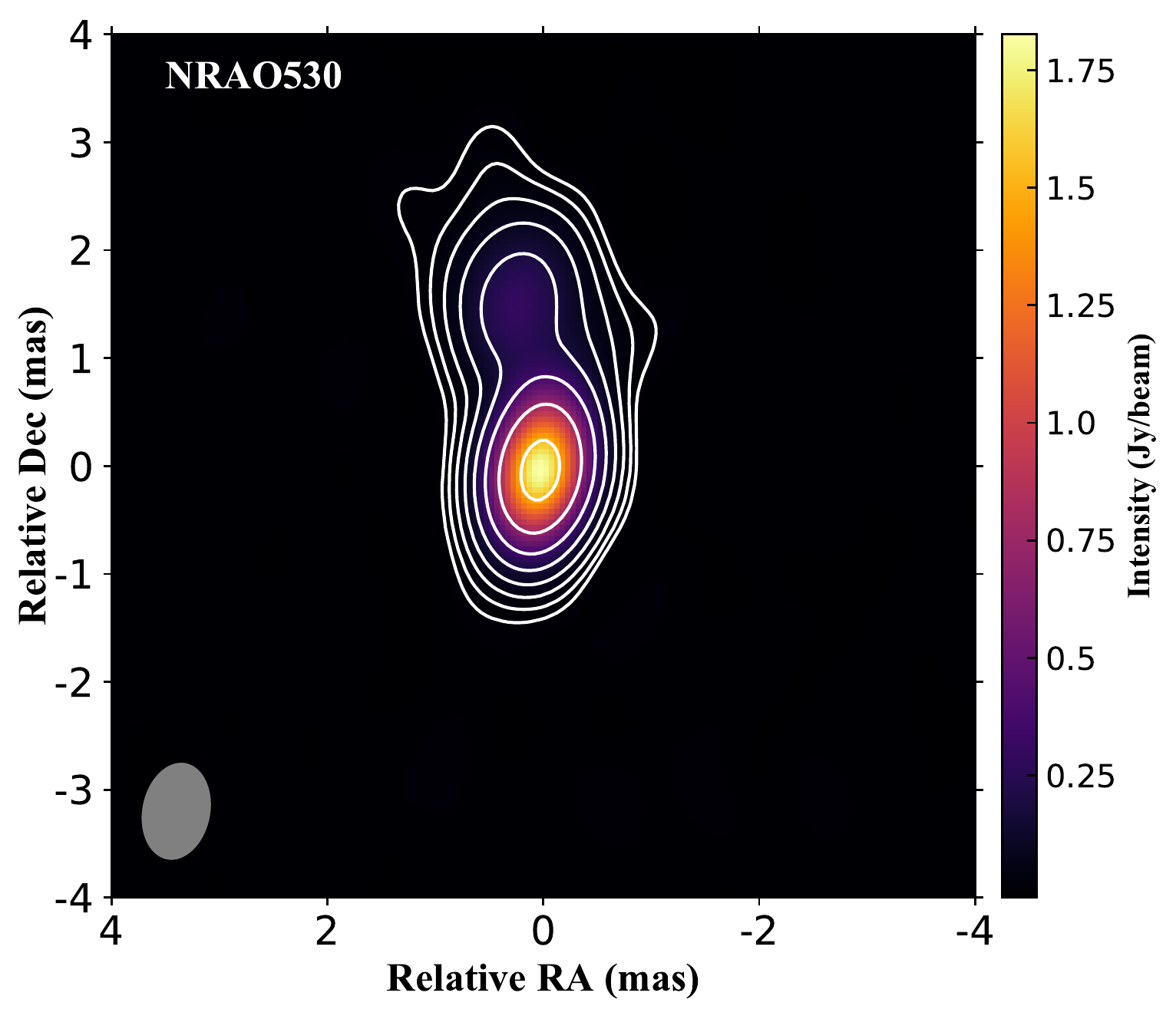}
    \includegraphics[height=5.5cm]{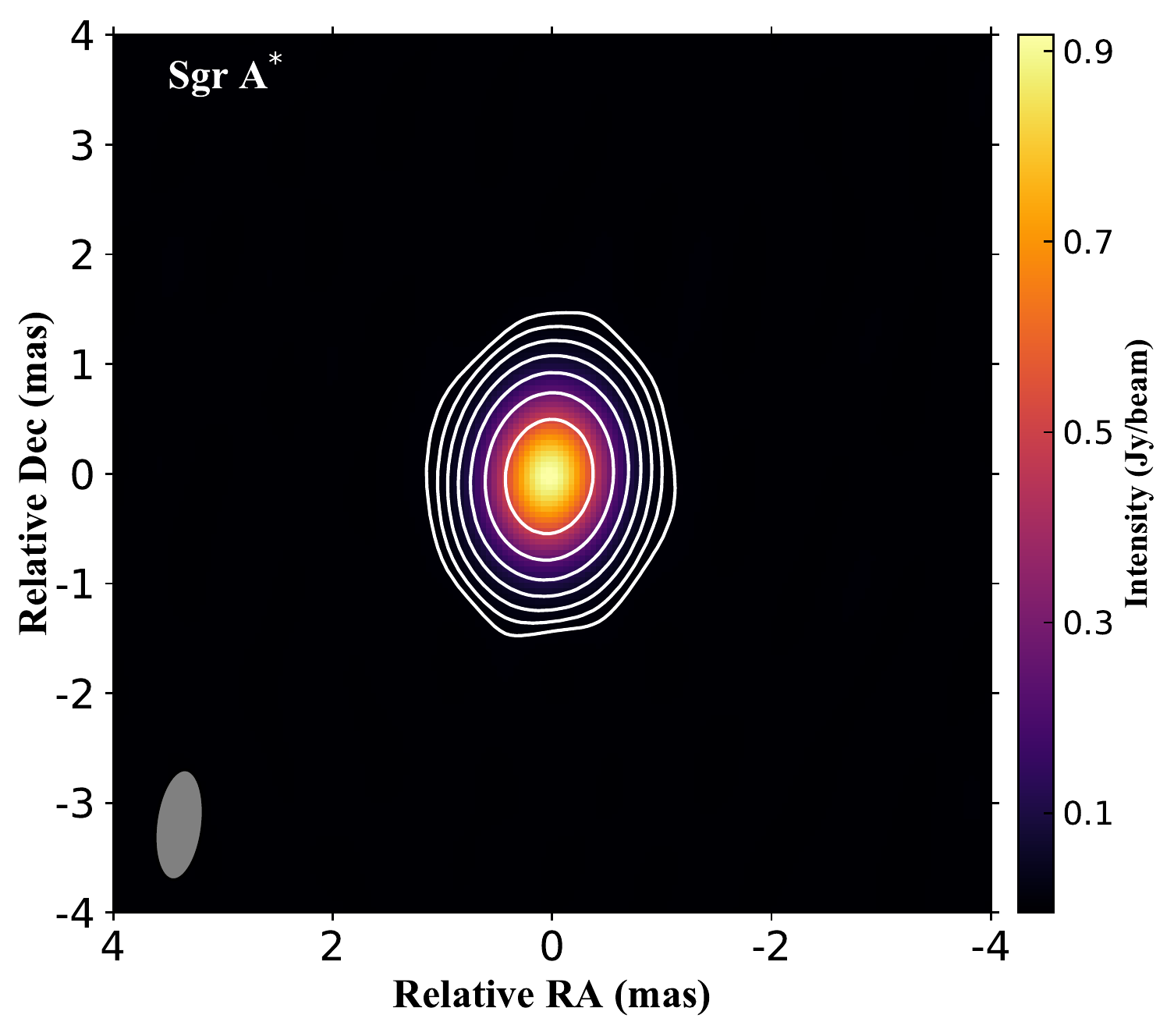}
    \\
    \caption{Clean image of NRAO530 (\textbf{left}) and model-fitted image of \sgra(\textbf{right}) on 11 May 2019 at 43 GHz. The~restoring beam is shown on the lower-left side of each panel. The~contour levels start at three times the rms noise value and positive levels increase by a factor of 2. The~image parameters are listed in Table~\ref{log}.}
    \label{fig:image}
\end{figure}

Figure~\ref{fig:radplot} shows the self-calibrated visibility amplitudes, as~a function of the ($u,v$)-distance. 
Since there are more than half of the data in the ``short'' baseline range, we can still get the gain corrections for all the antennas with only the data in this range (see Table~\ref{gain}). 
The resultant image of \sgra, from~the iterative Gaussian model fitting and self-calibration, is shown in Figure~\ref{fig:image} (right). 
Note that this is an ensemble-average image which is scatter-broadened, and~the derived structural parameters are listed in Table~\ref{modelfit} which are close to the size of the asymptotic Gaussian scattering kernel (e.g.,~\citep{2022ApJ...926..108C}). 
This also supports that the self-calibration with the Gaussian model fitting reasonably alternates the imaging for \sgra at these frequencies, as~the scattering kernel dominates the structure (see also~\citep{2022ApJ...926..108C} for comparison between model fitting and imaging). 
The derived station gains (Table~\ref{gain}) confirm that there is no significant biases in the models. 
Note that the image of NRAO530 from CLEAN imaging and the gain corrections are also shown in Figure~\ref{fig:image} (left) and Table~\ref{gain}, respectively, for~comparison.

\unskip

\unskip

\begin{table}[H]
\tablesize{\fontsize{7}{7}\selectfont}
\caption{Gaussian Model Fitting~Results.\label{modelfit}}
\centering
\begin{adjustwidth}{-\extralength}{0cm}
		\newcolumntype{C}{>{\centering\arraybackslash}X}
		\begin{tabularx}{\fulllength}{C cccccccccc}
\toprule
\textbf{Date} & 
\textbf{Method} & 
\textbf{S}\boldmath{$_{\nu}^{\rm tot}$} & 
\boldmath{$\theta_{\rm maj}^{\rm en}$}  & 
\boldmath{$\theta_{\rm min}^{\rm en}$}  & 
\boldmath{${r}^{\rm en}$} & 
\boldmath{$\theta_{\rm PA}^{\rm en}$}   &
\boldmath{$\theta_{\rm maj}^{\rm int}$} & 
\boldmath{$\theta_{\rm min}^{\rm int}$} & 
\boldmath{${r}^{\rm int}$} & 
\boldmath{$\theta_{\rm PA}^{\rm int}$}    \\
\textbf{(Exp.code)}    & 
        &
\textbf{(Jy)}    &
\textbf{(}\boldmath{$\upmu$}\textbf{as)}&
\textbf{(}\boldmath{$\upmu$}\textbf{as)}&
        &
\textbf{(deg)}   &
\textbf{(}\boldmath{$\upmu$}\textbf{as)}&
\textbf{(}\boldmath{$\upmu$}\textbf{as)}&
        &
\textbf{(deg)} \\
\textbf{(1)}     & 
\textbf{(2) }    &
\textbf{(3)}     & 
\textbf{(4)}    &
\textbf{(5)}     &
\textbf{(6)}     &
\textbf{(7)}     &
\textbf{(8)}     &
\textbf{(9)}     &
\textbf{(10)}    &
\textbf{(11)}  \\
\midrule
2019-02-27 & Gfit/Amp & 1.61$\,\pm\,$0.16 & 727.8$\,\pm\,$8.9 &  418.8$\,\pm\,$16.1 &  1.74$\,\pm\,$0.07 &  81.2$\,\pm\,$0.8 & 
293.6$\,\pm\,$23.9 &  238.4$\,\pm\,$34.8 &  1.23$\,\pm\,$0.21 &  73.9$\,\pm\,$7.2   \\
(a19mk01a) & CA & ... & 731.8$\,\pm\,$9.2 &  416.4$\,\pm\,$25.6 &  1.76$\,\pm\,$0.22 &  84.1$\,\pm\,$1.8 & 
303.8$\,\pm\,$24.7 &  229.5$\,\pm\,$60.8 &  1.32$\,\pm\,$0.60 &  92.5$\,\pm\,$13.1  \\
\midrule
2019-03-22 & Gfit/Amp & 1.48$\,\pm\,$0.15 & 737.2$\,\pm\,$8.6 &  412.0$\,\pm\,$11.5 &  1.79$\,\pm\,$0.05 &  82.6$\,\pm\,$0.5 & 
336.2$\,\pm\,$22.8 &  236.6$\,\pm\,$19.9 &  1.41$\,\pm\,$0.11 &  86.5$\,\pm\,$3.4   \\
(a19kh01a)& CA & ... & 739.4$\,\pm\,$9.2 &  421.7$\,\pm\,$20.8 &  1.75$\,\pm\,$0.08 &  82.1$\,\pm\,$1.2 & 
341.5$\,\pm\,$28.9 &  247.8$\,\pm\,$41.9 &  1.36$\,\pm\,$0.20 &  85.0$\,\pm\,$7.8  \\
\midrule
2019-03-29 & Gfit/Amp & 1.42$\,\pm\,$0.14 & 734.6$\,\pm\,$8.6 &  449.3$\,\pm\,$12.1 &  1.63$\,\pm\,$0.03 &  83.5$\,\pm\,$0.6 & 
317.1$\,\pm\,$25.6 &  283.9$\,\pm\,$11.2 &  1.13$\,\pm\,$0.04 &  114.1$\,\pm\,$12.1   \\
(a19kh01c) & CA & ... & 734.3$\,\pm\,$10.0 &  458.3$\,\pm\,$22.8 &  1.60$\,\pm\,$0.07 &  82.6$\,\pm\,$1.1 & 
317.8$\,\pm\,$25.0 &  298.9$\,\pm\,$20.5 &  1.05$\,\pm\,$0.07 &  122.7$\,\pm\,$42.0  \\
\midrule
2019-04-12 & Gfit/Amp & 1.88$\,\pm\,$0.19 & 736.2$\,\pm\,$8.7 &  483.6$\,\pm\,$10.4 &  1.52$\,\pm\,$0.03 &  85.5$\,\pm\,$0.6 & 
361.7$\,\pm\,$25.1 &  292.8$\,\pm\,$8.8 &  1.24$\,\pm\,$0.04 &  142.1$\,\pm\,$5.1   \\
(a19kh01f) & CA & ... & 737.0$\,\pm\,$9.1 &  482.0$\,\pm\,$32.6 &  1.53$\,\pm\,$0.11 &  86.2$\,\pm\,$1.6 & 
348.6$\,\pm\,$25.1 &  293.6$\,\pm\,$25.8 &  1.26$\,\pm\,$0.13 &  137.4$\,\pm\,$17.1  \\
\midrule
2019-05-11 & Gfit/Amp & 2.04$\,\pm\,$0.20 & 729.3$\,\pm\,$8.5 &  458.4$\,\pm\,$8.1 &  1.59$\,\pm\,$0.03 &  80.8$\,\pm\,$0.7 & 
322.2$\,\pm\,$23.5 &  300.2$\,\pm\,$9.2 &  1.08$\,\pm\,$0.04 &  17.2$\,\pm\,$22.4   \\
(a19mk01c) & CA & ... & 730.2$\,\pm\,$9.7 &  459.1$\,\pm\,$26.9 &  1.58$\,\pm\,$0.09 &  83.6$\,\pm\,$1.7 & 
312.6$\,\pm\,$27.5 &  289.9$\,\pm\,$15.8 &  1.08$\,\pm\,$0.11 &  130.7$\,\pm\,$37.9  \\
\midrule
2019-09-10 & Gfit/Amp & 1.67$\,\pm\,$0.17 & 732.6$\,\pm\,$8.9 &  467.4$\,\pm\,$16.5 &  1.57$\,\pm\,$0.04 &  81.1$\,\pm\,$0.7 & 
370.4$\,\pm\,$27.3 &  244.4$\,\pm\,$9.8 &  1.52$\,\pm\,$0.04 &  80.9$\,\pm\,$1.3   \\
(a19mk01e) & CA & ... & 735.5$\,\pm\,$8.8 &  471.0$\,\pm\,$23.6 &  1.57$\,\pm\,$0.10 &  82.8$\,\pm\,$7.7 & 
361.6$\,\pm\,$30.3 &  240.7$\,\pm\,$57.7 &  1.50$\,\pm\,$0.41 &  95.7$\,\pm\,$10.8  \\
\midrule
2019-10-10 & Gfit/Amp & 1.23$\,\pm\,$0.12 & 714.2$\,\pm\,$8.7 &  450.4$\,\pm\,$12.1 &  1.58$\,\pm\,$0.04 &  83.6$\,\pm\,$0.8 & 
298.7$\,\pm\,$23.9 &  248.0$\,\pm\,$10.5 &  1.21$\,\pm\,$0.08 &  151.1$\,\pm\,$7.7   \\
(a19mk01g) & CA & ... & 722.7$\,\pm\,$9.0 &  434.9$\,\pm\,$26.0 &  1.76$\,\pm\,$0.15 &  82.5$\,\pm\,$1.6 & 
310.0$\,\pm\,$26.0 &  262.3$\,\pm\,$20.6 &  1.15$\,\pm\,$0.22 &  94.0$\,\pm\,$39.7  \\
\midrule
2019-11-23 & Gfit/Amp & 1.71$\,\pm\,$0.17 & 738.2$\,\pm\,$8.7 &  412.8$\,\pm\,$12.4 &  1.79$\,\pm\,$0.06 &  81.1$\,\pm\,$0.5 & 
337.7$\,\pm\,$23.1 &  220.6$\,\pm\,$23.6 &  1.53$\,\pm\,$0.22 &  77.4$\,\pm\,$2.3   \\
(a19mk01h) & CA & ... & 735.7$\,\pm\,$9.2 &  417.2$\,\pm\,$22.1 &  1.76$\,\pm\,$0.14 &  80.9$\,\pm\,$1.1 & 
336.2$\,\pm\,$27.9 &  248.7$\,\pm\,$34.7 &  1.35$\,\pm\,$0.56 &  77.1$\,\pm\,$5.6  \\
\midrule
2019-12-18 & Gfit/Amp & 1.35$\,\pm\,$0.14 & 726.2$\,\pm\,$8.6 &  412.3$\,\pm\,$15.2 &  1.76$\,\pm\,$0.06 &  79.7$\,\pm\,$0.8 & 
295.9$\,\pm\,$24.3 &  222.3$\,\pm\,$26.6 &  1.33$\,\pm\,$0.17 &  61.3$\,\pm\,$7.7   \\
(a19mk01i) & CA & ... & 722.3$\,\pm\,$8.9 &  399.7$\,\pm\,$25.5 &  1.76$\,\pm\,$0.02 &  80.8$\,\pm\,$2.5 & 
281.8$\,\pm\,$26.6 &  214.8$\,\pm\,$62.9 &  1.33$\,\pm\,$0.49 &  77.9$\,\pm\,$23.2  \\
\bottomrule
\end{tabularx}
\end{adjustwidth}
\end{table}
\unskip

\section{Results}
\label{3}
\unskip
\subsection{Flux Density~Variability}

\begin{figure}[H]
    \includegraphics{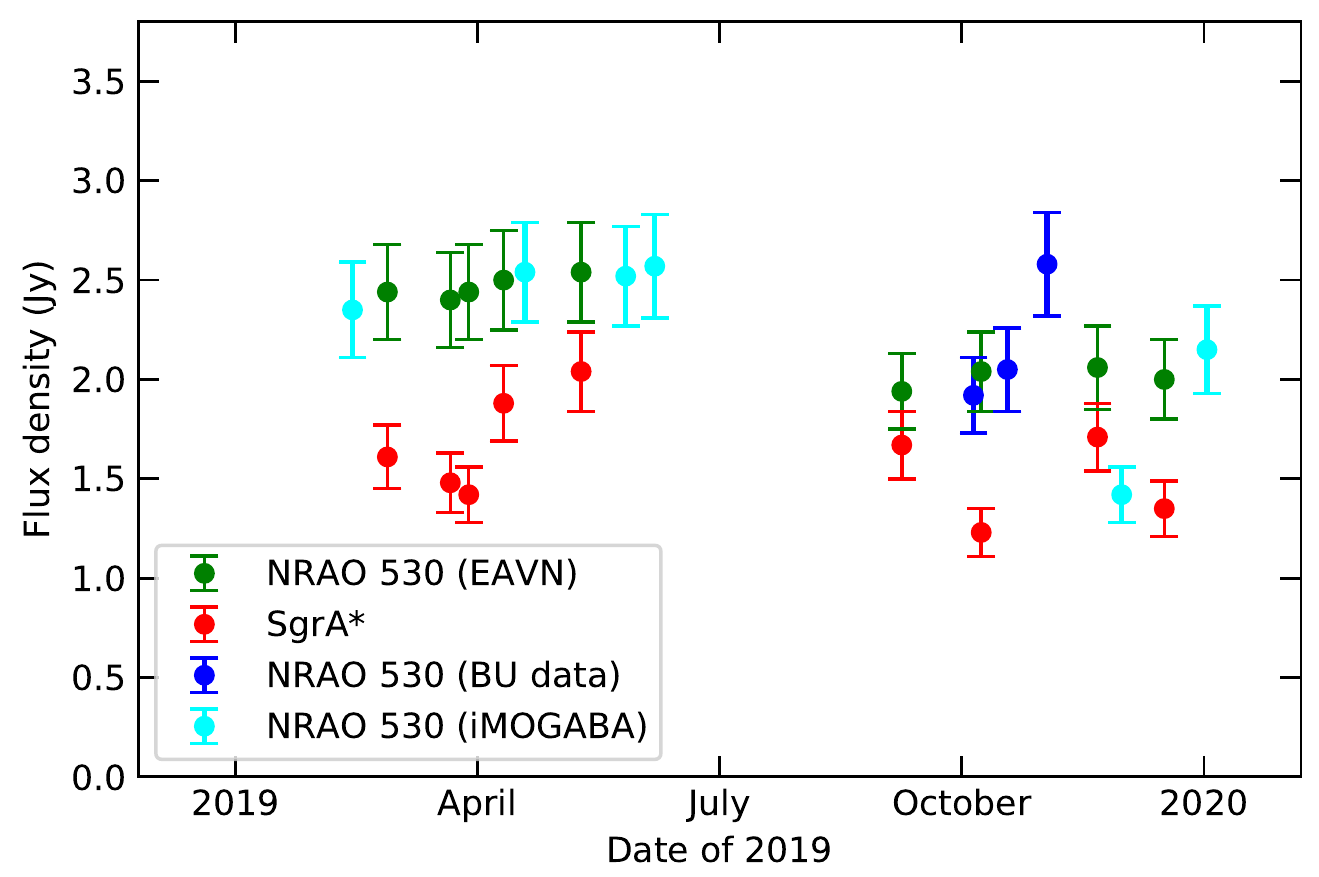}
    \caption{Light curve of \sgra and NRAO 530 at 43 GHz in 2019. Red points are the flux densities of \sgra at 43 GHz from our EAVN observations. Other points are flux densities of NRAO 530: green points are from our EAVN observations; blue points are from the BU observations in the second half of 2019; cyan points are from the iMOGABA observations. The~error bars are 10\% of the total flux~density.}
    \label{fig:LC}
\end{figure}

Total flux density of \sgra, S$\rm_{tot}$, is found from the final fitted Gaussian model, together with the parameters of elliptical structure (i.e., $\theta_{\rm maj}^{\rm en}$, ${\rm r}^{\rm en}$, and~${\rm PA}^{\rm en}$, which are major axis size, axial ratio, and~position angle of the major axis, respectively; see Table~\ref{modelfit}).
In Figure~\ref{fig:LC}, we show the light curve of \sgra at 43 GHz in 2019. 
The mean flux density is about 1.60\,Jy, which is higher than the previous measurements with EAVN in April 2017 ($\sim$1.36~$\pm$~0.14~Jy;~\citep{2022ApJ...926..108C}) and with VERA in November 2004--April 2009 (0.9~$\pm$~0.1 Jy;~\citep{2013PASJ...65...91A}) but slightly lower than the VLBA results in 2007 (1.79~$\pm$~0.05 Jy;~\citep{2011A&A...525A..76L}). 
The flux density of \sgra appear more pronounced in April and May, during~a time that coincides with two detected NIR flares occurring on 20 April and 13 May 2019.
The highest flux density detected in our observations is 2.04~$\pm$~0.20 Jy on 11 May 2019.

The amount of variability has been estimated with the variability index, $m=\sigma$/<$S$>, where the $\sigma$ and <$S$> are the standard deviation and mean flux density, respectively. 
As a result, the~$m$ of \sgra and NRAO 530 are obtained as \xc{15.4$\pm$3.1\% and $\sim$2\%}, respectively. 
This clearly shows the source intrinsic flux variation of \sgra, which is remarkably larger than the previous measurements, for~instance in May 2007 ($m=9.3$\%;~\citep{2011A&A...525A..76L}) and between 2004 and 2008 ($\sim$11\%;~\citep{2013PASJ...65...91A}). 
Note that the $m$ of NRAO\,530 has been estimated for the first half (\xc{$m=1.9\pm0.5\%$}) and second half (\xc{$m=2.1\pm0.5\%$}) of 2019 separately. 
This is because the light curve of NRAO 530 shows progressive decrease over the year, which is also shown from independent observations with the VLBA-BU-BLAZAR monitoring program~\citep{Jorstad_2017, Weaver_2022} and the iMOGABA program with the KVN~\citep{Lee_2016} (see Figure~\ref{fig:LC}). 
Note that the VLBA-BU-BLAZAR data are only used after September 2019, due to pointing problems at mm wavelengths \endnote{VLBA Test Memos \#\,73. VME Transition VLBA Pointing Issues (\url{https://library.nrao.edu/public/memos/vlba/test/VLBAT_73.pdf}, (accessed on 1 November 2022)); 
VLBA-BU-BLAZAR program (\url{https://www.bu.edu/blazars/BEAM-ME.html}, (accessed on 1 November 2022)).}

\subsection{Intrinsic structure and its correlation with the flux~density}

To obtain the intrinsic structure of \sgra, the~self-calibrated visibilities and closure amplitudes are deblurred by using the recent scattering kernel model (see \citet{2022ApJ...926..108C} for more details).
Then an elliptical Gaussian model with three free parameters ($\theta_{\rm maj}^{\rm int}$, ${\rm r}^{\rm int}$, and~${\rm PA}^{\rm int}$) is fitted to the deblurred log closure amplitudes (CA) and visibility amplitudes (Gfit/Amp), using the Monte Carlo method (Table \ref{modelfit}). 
As a result, the mean values of $\theta_{\rm maj}^{\rm int}$ and $\theta_{\rm min}^{\rm int}$ are found as $\sim$324 and 258 $\upmu$as, respectively, which are slightly larger than the previous study~\citep{2022ApJ...926..108C}, with the mean position angle (PA) roughly consistent with the previous studies~\citep{2007MNRAS.379.1519M,2014ApJ...790....1B,2022ApJ...926..108C}. 
Note, however, that the intrinsic PA is less constrained, mainly because the intrinsic shape is close to circular, as~shown in the large deviation of PA between two different methods (Gfit/Amp and CA) and across the observations. 
The uncertainties are estimated by the goodness-of-fit from the Monte Carlo method and the stochastic random phase screen within the error range of scattering parameters (see \citet{2022ApJ...926..108C} for more details), so that the final uncertainties are the combination of them. 
For the latter one, the~scattering screen provides the refractive noise and the error range of the size of scattering kernel. 
As for the refractive noise, it is used to flag the noise-dominated data based on the S/N (see Section~\ref{2.3}), not directly added to the visibilities. 
The scattering kernel size is mostly determined by the power-law index of the phase structure function of the scattering screen, $\alpha$, and~$r_{\rm in}$ (e.g.,~\citep{2018arXiv180501242P, 2018ApJ...865..104J}). 
Since the intrinsic sizes are obtained from deblurred data (i.e., division by the scattering kernel in the visibility domain), the~uncertainties of the kernel size which are introduced by the error range of $\alpha$ and $r_{\rm in}$ are used for the error of intrinsic sizes as a quadratic sum, together with the fitting error.

Similar to the flux variation, remarkably, the~intrinsic sizes also show the variation across our observations. 
The variability index, $m$, is obtained as $6.8\pm1.5\%$ and $10.3\pm1.7\%$ for major and minor axis sizes, respectively. 
Note that the $m$ of observed (i.e., scattered) sizes are $0.8\pm0.1\%$ and $5.7\pm0.9\%$ for each axis. 
These are consistent with the previous studies~\citep{2011A&A...525A..76L, 2013PASJ...65...91A}, indicating marginal size variation towards the minor axis. 
Note, however, that the observed structure is dominated by the scattering kernel with PA$\sim$82$^\circ$--86$^\circ$ (e.g.,~\citep{2022ApJ...926..108C}) so that it is relatively easier to detect the size variation towards minor axis (i.e., north-south direction) when the intrinsic size is varying.   
On the other hand, the~$m$ of the area of intrinsic structure (i.e., $\pi$~$\times$~$\theta \rm ^{int}_{maj}$~$\times$~$\theta \rm^{int}_{min}$) is obtained as $13.3\pm3.5\%$, which is comparable to the flux variability ($\sim$15\%). 
These results suggest that the total flux density variations might be associated with changes in size and/or area.

\begin{figure}[H]
\includegraphics[width=\linewidth]{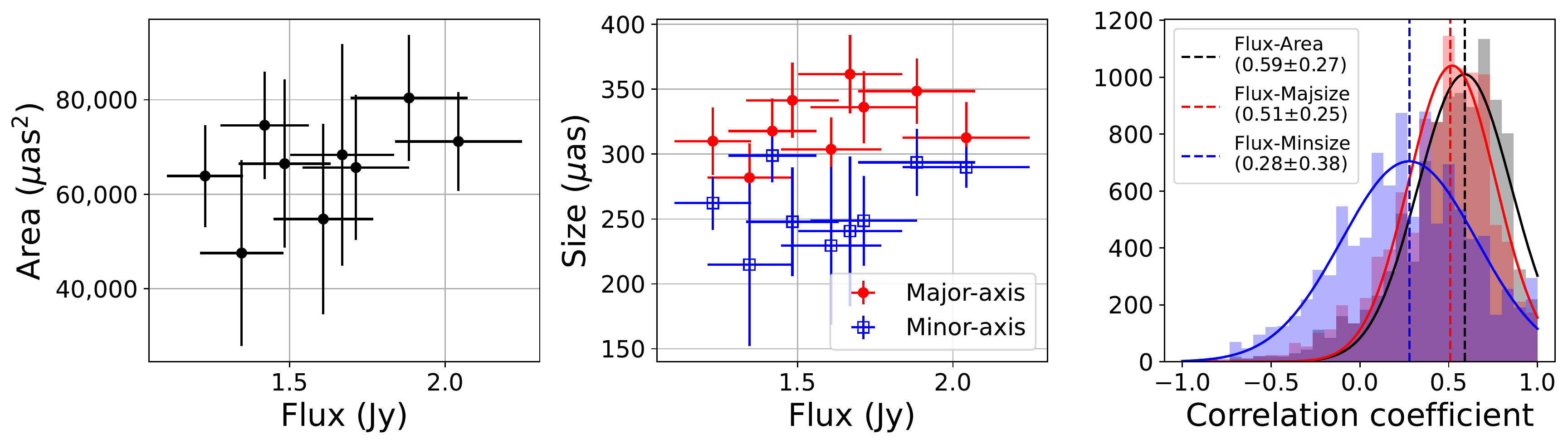}
\includegraphics[width=0.9\linewidth]{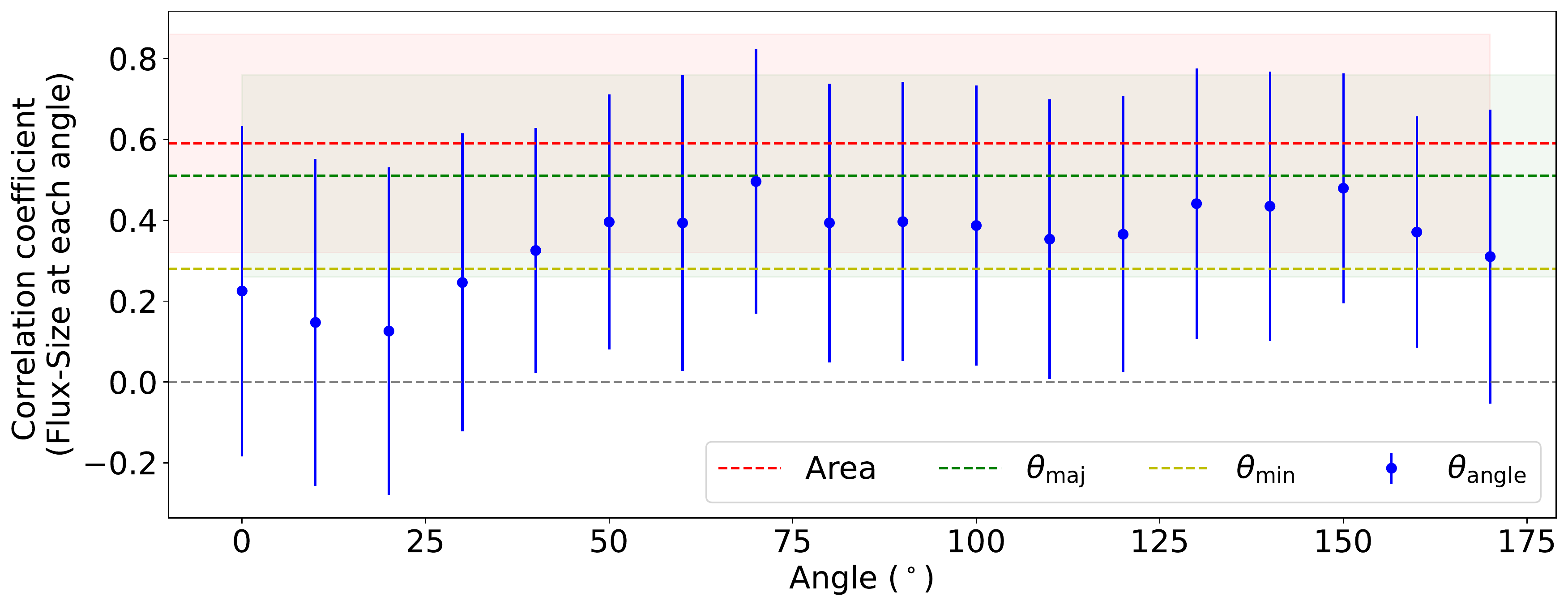}
\caption{
{(\textbf{Top}}): 
The correlation of flux density and area (i.e., $\pi\theta_{\rm maj}\theta_{\rm min}$/4; \textbf{left}), $\theta_{\rm maj}$ (\textbf{middle}, red), $\theta_{\rm min}$ (\textbf{middle}, blue) from our observations in 2019, where the $\theta_{\rm maj}$ and $\theta_{\rm min}$ are the major-axis and minor-axis size, respectively. 
Resampling distributions of the estimated correlation coefficients of each are also shown (right). (Bottom}): The correlation coefficients between flux density and size at each angle (east of north; with a step of $10^\circ$). No significant correlation is found that is larger than the ones from the area and major-axis size. 

\label{fig:correlation}
\end{figure}

While the previous studies have found a strong correlation between flux density and intrinsic minor axis size at 22, 43, and~86\,GHz, the~correlation with intrinsic major axis size has only been shown at 86\,GHz~\citep{2011A&A...525A..76L}. 
In order to further explore the possible association in our observations, 
we have estimated the Spearman rank correlation coefficient of the flux density versus area, major-axis, and~minor-axis sizes.
Since the number of data points is insufficient to get a reliable $p$-value, all data points are randomly resampled 10,000 times to obtain the reasonable uncertainty of correlation coefficient (resampling or bootstrapping method; e.g.,~\citep{curran_2014}). This considers the error of sample distribution, compared to the complete population distribution. 
In Figure~\ref{fig:correlation} (top), we show the correlation between flux and area (left), and~size of both major and minor axis (middle). The~distribution of the correlation coefficients from the resampling process is shown in the right panel. 
By this method, we have found the (marginal) correlation of flux density versus area and major-axis size with $\gtrsim$$2\,\sigma$, while no clear correlation has been found with minor-axis size. 
This provides different results from \citet{2011A&A...525A..76L}. 
Note, however, that the intrinsic structure of \sgra is close to circular Gaussian (i.e., the~axial ratio is close to unity) so that the PA is not well determined with large uncertainties (e.g., $\sim$60$\degree$ to $\sim$150$\degree$), unlike the observed structure. 
In this regard, the~variability of the size of each axis does not provide consistent directional information. 
To make it clearer, we have also estimated the correlation of the flux density with the size at different angles, but it is still difficult to find a preferred direction (Figure~\ref{fig:correlation}, bottom). 

\unskip

\section{Discussion}
\label{4}

In the previous section, we showed the correlation between the flux density and the emitting region area in 2019.
While recent observations of the innermost part of Sgr A* suggest 
that it is in a magnetically arrested disk (MAD) state 
with a smaller viewing angle (EHT and Gravity collaboration),
the cause of its variability is still a matter of debate.
While rapid variability on a timescale of 10 min in \sgra has been discussed 
and interpreted in the context of a transient feature (such as a hotspot) that appeared around the central black hole (e.g., \citet{2022A&A...665L...6W} and references therein),
the origin of flux variability on a monthly timescale is not yet well explored in \sgra.
Below, we discuss possible origins for the correlation seen in~2019.

The first possibility discussed here is the enhancement of the flux and emission area through (i) the emergence of a new outflow (jet), or~(ii) an increase in the size of the optically-thick region of the accretion flow.
It is well known that theoretical models of the dominating component of the radio emission in \sgra generally fall into two broad classes.
One is a weak and compact jet model, while the other is a radiatively inefficient accretion flow~\citep{1995Natur.374..623N,2000A&A...362..113F,2000ApJ...541..234O,2003ApJ...598..301Y}. 

Regarding case (I), in spite of intensive VLBI observations (e.g.,~\citep{1993A&A...277L...1A,1999A&A...343..801M,2004Sci...304..704B,2005Natur.438...62S,2014ApJ...790....1B}), there is no clear evidence of a jet-like structure yet, although there are some indirect suggestions of the possible existence of a weak outflow/jet (e.g.,~\citep{2022ApJ...930L..12E,2022ApJ...930L..13E,2022ApJ...930L..14E,2022ApJ...930L..15E,2022ApJ...930L..16E,2022ApJ...930L..17E}). 
When the weak jet was ejected in \sgra, especially towards the line of sight,
it naturally explains the correlation between the flux density and the emitting region area in 2019.
In case (ii), a~temporary increase in the mass accretion rate may result in the expansion of the optically-thick region of the accretion flow.
An increase of the mass accretion rate
can alter not only a change in the number density of the electrons in the accretion flow, but~also the electron temperature and magnetic field in the accretion flow. 
Therefore, proper determination of the corresponding increase in the flux and the size of the emission region requires a dedicated study. Although~it is beyond the scope of this paper.
The need for such theoretical studies is expected to grow in the future.

\begin{figure}[H]
\includegraphics[width=0.9\linewidth]{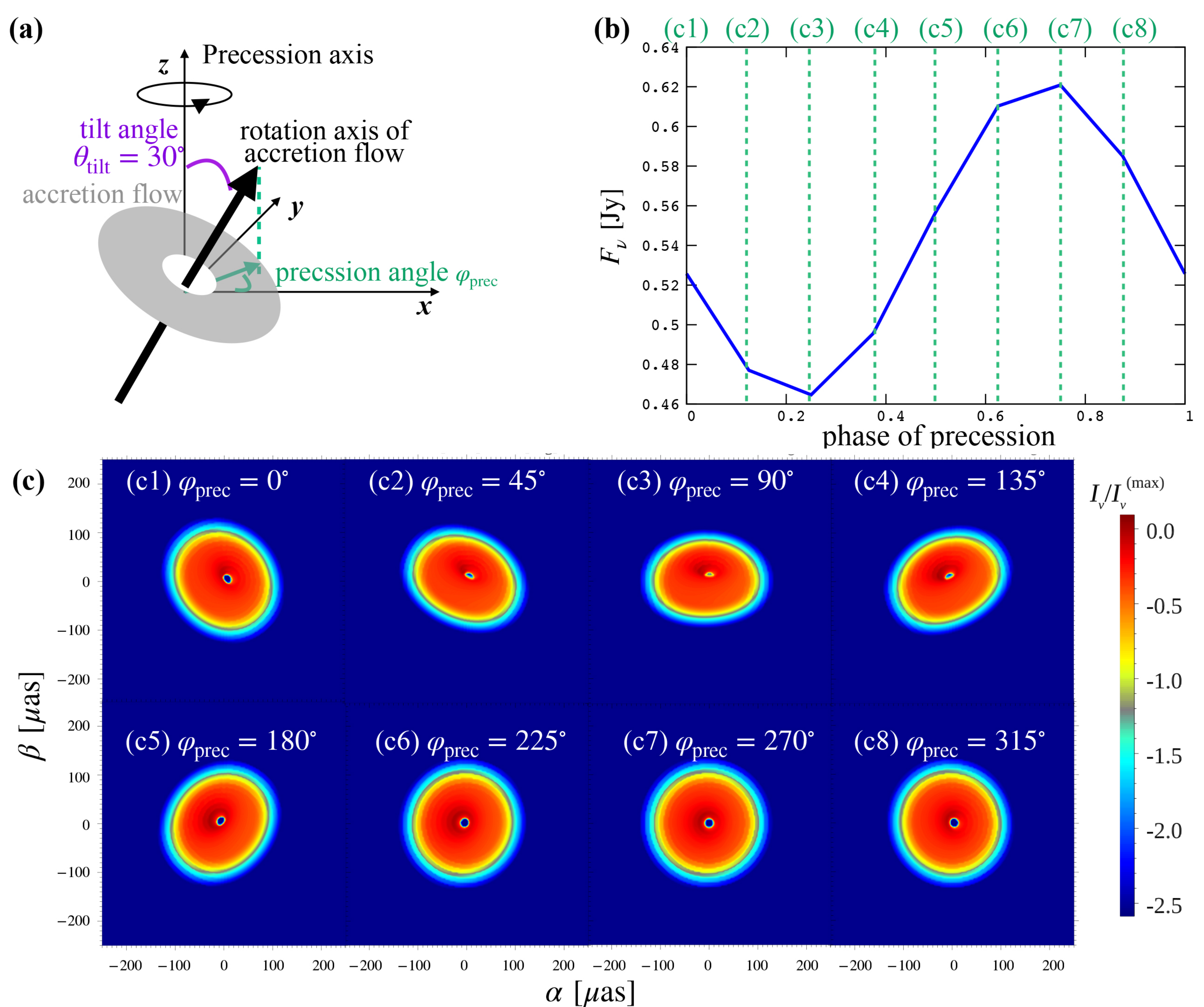}
\caption{(a): Schematic picture of a precessing accretion flow. (b) Light curve. (\textbf{c}) GRRT images of precessing accretion flow. The~tilt angle is set to be $\theta_{\rm tilt} = 30^{\circ}$. The~precession angle $\varphi_{\rm prec}$ is set at every $45^{\circ}$ from $0^{\circ}$ to $315^{\circ}$. Here, the~spherical coordinate is defined as $\cos \theta \equiv z/\sqrt{x^2 + y^2 + z^2}$,  $\cos \phi \equiv x/\sqrt{x^2 + y^2}$ and  $\sin \phi \equiv y/\sqrt{x^2 + y^2}$ by using the Cartesian coordinate depicted in the panel (\textbf{a}). The~observer locates in the direction $(\theta_{\rm ob}, \varphi_{\rm ob}) = (30^{\circ}, 270^{\circ})$. We note that $\theta_{\rm ob}$ is the angle between the precession axis (not the rotation axis of accretion flow) and the line of~sight.}

\label{fig:GRRTimage}
\end{figure}

The second possibility discussed here is the change of the accretion disk axis.
Recent general relativistic magnetohydrodynamic (GRMHD) simulation of a wind-fed accretion model for \sgra proposed by \citet{2020ApJ...896L...6R}
suggests that the accretion disk is tilted with the varying angle from $\sim$20--30$^{\circ}$ to $0^{\circ}$, with respect to the initial angular momentum axis for the model of stellar-winds injection with lower plasma $\beta$.
\endnote{The tilt angle can achieve $\sim 90^{\circ}$ for the model of stellar-winds injection with higher plasma $\beta$ and sometimes, for~the lower $\beta$ model.}
The origin of the tilt can be explained by the change of the  direction of the net magnetic flux conveyed onto the central black hole.
The rotation axis of the accretion flow can change both due to the variation of the tilt angle itself and the subsequent precession
of accretion flow induced by the tilt against the black-hole spin axis, i.e.,~the Lense-Thirring precession.
\endnote{
{The Figure~C1 of \citet{2020ApJ...896L...6R}
shows significant variations 
of the orientation of the angle-averaged angular momentum  
with respect to the angle-averaged magnetic field.}
 }
Observationally, EAVN at 22 and 43 GHz for \sgra show that the axial ratio of the intrinsic major-axis size to the intrinsic minor-axis size 
is about $26/20=1.3$. The~viewing angle of the accretion disk,
defined as the angle between the rotation axis of accretion flow and the line of sight (see Figure~\ref{fig:GRRTimage}), is consistent with  $\theta_{\rm view}\sim$$30^{\circ}$~\citep{2022ApJ...926..108C}.
Therefore, the~wind-fed model
and/or the precession model can realize the decrease of $\theta_{\rm view}$ that explains the correlation between the flux density and the emitting region area. 

In Figure~\ref{fig:GRRTimage}, we demonstrate an example of the precessing Kepleriian accretion flow model. 
For simplicity, only thermal electrons are considered in the accretion flow, and~the fixed tilt angle of $\theta_{\rm tilt} = 30^{\circ}$ is assumed.
We set the observation angle, which is the angle between the precession axis and the line of sight, to~be $\theta_{\rm ob} = 30^{\circ}$. 
The viewing angle can change as a consequence of the precession.
The images and the light curves at  $\nu = 43$~GHz are computed by solving general relativistic radiative transfer (GRRT) equations by using \texttt{RAIKOU} code~\cite{2019ApJ...878...27K, arXiv:2108.05131}.
The peak luminosity appears at $\varphi_{\rm prec} = 270^{\circ}$, corresponding to 
the face-on view of the accretion flow.
This is because the accretion flow is optically thick against the synchrotron-self-absorption at 43 GHz.
The predicted light curve shows $\sim$ a 
10\% variability, which is slightly less remarkable than 
the EAVN observation data. 
An inclusion of nonthermal electrons would mitigate 
the discrepancy between the  model and the observation.
For instance,
prominent eruptions of the plasma at the interface of the jet and the precessing accretion flow are indicated in GRMHD simulations 
that may induce the anisotropic injection of the nonthermal electrons.
The study of more sophisticated models remains as future~work.

\section{Summary}
\label{5}

In this study, we have found the temporal variability of flux density and intrinsic size of \sgra at 43\,GHz. 
The mean flux density is about 1.60\,Jy, which is larger than previous measurements that can 
be related to the NIR flare in 13 May 2019. Note that the highest flux density reaches 2.04 $\pm$ 0.20 Jy on 11 May 2019, from our observations. 
Variability is quantified with the $m-$index, which is provided as \xc{15.4\% and 2.1\%} for \sgra and NRAO\,530, respectively, indicating the large source intrinsic variability of \sgra. 
In addition to the flux variation, we also investigate the variation of intrinsic size and their possible correlations. 
After the scattering mitigation, we have first derived the intrinsic structure of \sgra, which can be well described with a single elliptical Gaussian model. 
As a result, across 9 epochs of observations in 2019, we have found a large variability of intrinsic size ($\sim$7\% and $\sim$10\% for major and minor axis size, respectively). 
With this, marginal ($\gtrsim$2\,$\sigma$) correlations of flux density versus area and major axis size are also found.

Two possible scenarios have been considered to explain the variability and the correlation of flux density and intrinsic size. 
First, a~weak jet towards the line of sight or the increased size of the optically-thick region of an accretion flow. 
Both can explain the observed variability and the correlation, especially if they are induced by the NIR flare. 
As for the jet scenario, for~instance, \citet{Rauch_2016} have found a secondary VLBI component at 43\,GHz, which may be triggered by the NIR flare, although~the possibility of refractive sub-structure cannot be fully ruled out. 
This also indicates that the detectability of the possible episodic jet increases through the multi-wavelength (MWL) (quasi-)simultaneous observations, as~well as the amount of the NIR activity, so that the MWL associated VLBI monitoring observations are of great importance for future detection. 
The next possible scenario is the tilted disk, based on a wind-fed accretion model, so that the rotation axis of the accretion disk shows precession. 
This results in the variability of both structure (i.e., area and size towards each axis) and flux density (e.g., projected area towards line of sight is changed), so the scenario also well explains our observations. 
With this, the~NIR flare may be related to significant changes of tilt angle, but it needs further investigation.
To prove the scenario, in~addition, better constraints on the intrinsic PA through tighter scattering parameter constraints (G.-Y.~Zhao~et~al. in prep.; Y.~Kofuji~et~al. in prep.) will be crucial.

\vspace{6pt} 



\authorcontributions{X.C. analyzed the VLBI data. X.C., I.C., T.K., and~M.K. wrote the original manuscript. X.C. and I.C. contributed to the data analysis. All the authors contribute to the discussion of the result of this work. All authors have read and agreed to the published version of the~manuscript.}

\funding{
The work at the IAA-CSIC is supported in part by the Spanish Ministerio de Econom{\'i}a y Competitividad (grants AYA2016-80889-P and PID2019-108995GB-C21), the~Consejería de Econom{\'i}a, Conocimiento, Empresas y Universidad of the Junta de Andaluc{\'i}a (grant P18-FR-1769), the~Consejo
Superior de Investigaciones Cient{\'i}ficas (grant 2019AEP112), and~the State Agency for Research of the Spanish MCIU through the ``Center of Excellence Severo Ochoa'' award to the Instituto de Astrof{\'i}sica de Andaluc{\'i}a (SEV-2017-0709). 
This work was also supported by JSPS KAKENHI Grant Numbers JP18K13594, and~MEXT as ``Program for Promoting Researches on the Supercomputer Fugaku'' (Toward a unified view of the universe: from large scale structures to planets, JPMXP1020200109) (T.K). 
Numerical computations were in part carried out on Cray XC50 at Center for Computational Astrophysics, National Astronomical Observatory of Japan.
This work was supported by Brain Pool Program through the National Research Foundation of Korea (NRF) funded by the Ministry of Science and ICT (2019H1D3A1A01102564). }

\acknowledgments{
This work is made use of the East Asian VLBI Network (EAVN), which is operated under cooperative agreement by National Astronomical Observatory of Japan (NAOJ), Korea Astronomy and Space Science Institute (KASI), Shanghai Astronomical Observatory (SHAO). We acknowledge all staff members and students who supported the operation of the array and the correlation of the data. This study partially makes use of 43\,GHz VLBA data from the VLBA-BU Blazar Monitoring Program (VLBA-BU-BLAZAR; \url{http://www.bu.edu/blazars/VLBAproject.html}), funded by NASA through the Fermi Guest Investigator grant 80NSSC20K1567.
}




\dataavailability{All observing data obtained by KaVA or EAVN, except the data in the term of the right of occupation by a principal investigator, are archived via the following website. \url{https://radio.kasi.re.kr/arch/search.php}.} 


\conflictsofinterest{The authors declare no conflict of~interest.} 




%

\appendixtitles{no} 

\begin{adjustwidth}{-\extralength}{0cm}

\printendnotes[custom]
\reftitle{References}

\PublishersNote{}
\end{adjustwidth}

\begin{thebibliography}{999}
\bibitem[Genzel(2010)]{genzel2010} Genzel, R.; Eisenhauer, F.; Gillessen, S. The Galactic Center massive black hole and nuclear star cluster. {\em Rev. Mod. Phys.} {\bf 2010}, {\em 82}, 3121--3195.
\bibitem[GRAVITY(2020)]{gravity2020} Abuter, R.~et~al. [GRAVITY Collaboration] Detection of the Schwarzschild precession in the orbit of the star S2 near the Galactic centre massive black hole. {\em Astron. Astrophys.} {\bf 2020}, {\em 636}, L5.
\bibitem[GRAVITY Collaboration et al.(2019)]{gravity2019} Abuter, R.~et~al. [GRAVITY Collaboration] A geometric distance measurement to the Galactic center black hole with 0.3\% uncertainty. {\em Astron. Astrophys.} {\bf 2019}, {\em 625}, L10.
\bibitem[Event Horizon Telescope Collaboration et al.(2022)]{2022ApJ...930L..12E} Akiyama, K.~et~al. [Event Horizon Telescope Collaboration]\ First Sagittarius A* Event Horizon Telescope Results. I. The Shadow of the Supermassive Black Hole in the Center of the Milky Way. {\em Astrophys. J. Lett.} {\bf 2022}, {\em 930}, L12.
\bibitem[Event Horizon Telescope Collaboration et al.(2022)]{2022ApJ...930L..13E} Akiyama, K.~et~al. [Event Horizon Telescope Collaboration]\ First Sagittarius A* Event Horizon Telescope Results. II. EHT and Multiwavelength Observations, Data Processing, and Calibration. {\em Astrophys. J. Lett.} {\bf 2022}, {\em 930}, L13.
\bibitem[Event Horizon Telescope Collaboration et al.(2022)]{2022ApJ...930L..14E} Akiyama, K.~et~al. [Event Horizon Telescope Collaboration]\ First Sagittarius A* Event Horizon Telescope Results. III. Imaging of the Galactic Center Supermassive Black Hole. {\em Astrophys. J. Lett.} {\bf 2022}, {\em 930}, L14.
\bibitem[Event Horizon Telescope Collaboration et al.(2022)]{2022ApJ...930L..15E} Akiyama, K.~et~al. [Event Horizon Telescope Collaboration]\ First Sagittarius A* Event Horizon Telescope Results. IV. Variability, Morphology, and Black Hole Mass. {\em Astrophys. J. Lett.} {\bf 2022}, {\em 930}, L15.
\bibitem[Event Horizon Telescope Collaboration et al.(2022)]{2022ApJ...930L..16E} Akiyama, K.~et~al. [Event Horizon Telescope Collaboration]\ First Sagittarius A* Event Horizon Telescope Results. V. Testing Astrophysical Models of the Galactic Center Black Hole. {\em Asgtrophys. J. Lett.} {\bf 2022}, {\em 930}, L16.
\bibitem[Event Horizon Telescope Collaboration et al.(2022)]{2022ApJ...930L..17E} Akiyama, K.~et~al. [Event Horizon Telescope Collaboration]\ First Sagittarius A* Event Horizon Telescope Results. VI. Testing the Black Hole Metric. {\em Astrophys. J. Lett.} {\bf 2022}, {\em 930}, L17.
\bibitem[Falcke \& Markoff(2000)]{2000A&A...362..113F} Falcke, H.; Markoff, S. The jet model for Sgr A*: Radio and X-ray spectrum. {\em Astron. Astrophys.} {\bf 2000}, {\em 362}, 113
\bibitem[Yuan et al.(2003)]{2003ApJ...598..301Y} Yuan, F.; Quataert, E.; Narayan, R. Nonthermal Electrons in Radiatively Inefficient Accretion Flow Models of Sagittarius A*. {\em Astrophys. J.} {\bf 2003}, {\em 598}, 301.
\bibitem[Narayan(1992)]{1992RSPTA.341..151N} Narayan, R. The Physics of Pulsar Scintillation. {\em Philos. Trans. R. Soc. Lond. Ser.} {\bf 1992}, {\em 341}, 151.
\bibitem[Bower et al.(2006)]{2006ApJ...648L.127B} Bower, G.C.; Goss, W.M.; Falcke, H.; Backer, D.C.; Lithwick, Y. The Intrinsic Size of Sagittarius A* from 0.35 to 6 cm. {\em Astrophys. J.} {\bf 2006}, {\em 648}, L127.
\bibitem[Bower et al.(2004)]{2004Sci...304..704B} Bower, G.C.; Falcke, H.; Herrnstein, R.M.; Zhao, Jun-Hui, Goss, W.M.; Backer, Donald, C. Detection of the Intrinsic Size of Sagittarius A* Through Closure Amplitude Imaging. {\em Science} {\bf 2004}, {\em 304}, 704.
\bibitem[Genzel et al.(2003)]{2003Natur.425..934G} Genzel, R.; Sch{\"o}del, R.; Ott, T.; Eckart, A.; Alexander, T.; Lacombe, F.; Rouan, D.; Aschenbach, B. Near-infrared flares from accreting gas around the supermassive black hole at the Galactic Centre. {\em Nature} {\bf 2003}, {\em 425}, 934.
\bibitem[Neilsen et al.(2013)]{2013ApJ...774...42N} Neilsen, J.; Nowak, M.A.; Gammie, C.; Dexter, J.; Markoff, S.; Haggard, D.; Nayakshin, S.; Wang, Q.D.; Grosso, N.; Porquet, D.; et~al. A Chandra/HETGS Census of X-Ray Variability from Sgr A* during 2012. {\em Astrophys. J.} {\bf 2013}, {\em 774}, 42.
\bibitem[Eckart et al.(2006)]{2006A&A...450..535E} Eckart, A.; Baganoff, F.K.; Sch{\"o}del, R.; Morris, M.; Genzel, R.; Bower, G.C.; Marrone, D.; Moran, J.M.; Viehmann, T.; Bautz, M.W.; et al. The flare activity of Sagittarius A*. New coordinated mm to X-ray observations. {\em Astron. Astrophys.} {\bf 2006}, {\em 450}, 535.
\bibitem[Yusef-Zadeh et al.(2012)]{2012ApJ...758L..11Y} Yusef-Zadeh, F.; Arendt, R.; Bushouse, H.; Cotton, W.; Haggard, D.; Pound, M.W.; Roberts, D.A.; Royster, M.; Wardle, M. A 3 pc Scale Jet-driven Outflow from \sgra. {\em Astrophys. J.} {\bf 2012}, {\em 758}, L11.
\bibitem[Ponti et al.(2017)]{2017MNRAS.468.2447P} Ponti, G.; George, E.; Scaringi, S.; Zhang, S.; Jin, C.; Dexter, J.; Terrier, R.; Clavel, M.; Degenaar, N.; Eisenhauer, F.; et al. A powerful flare from Sgr A* confirms the synchrotron nature of the X-ray emission. {\em Mon. Not. R. Astron. Soc.} {\bf 2017}, {\em 468}, 2447.
\bibitem[Fazio et al.(2018)]{2018ApJ...864...58F} Fazio, G.G.; Hora, J.L.; Witzel, G.; Willner, S.P.; Ashby, M.L.N.; Baganoff, F.; Becklin, E.; Carey, S.; Haggard, D.; Gammie, C.; et al. Multiwavelength Light Curves of Two Remarkable Sagittarius A* Flares. {\em Astrophys. J.} {\bf 2018}, {\em 864}, 58.
\bibitem[Marrone et al.(2008)]{2008ApJ...682..373M} Marrone, D.P.; Baganoff, F.K.; Morris, M.R.; Moran, J.M.; Ghez, A.M.; Hornstein, S.D.; Dowell, C.D.; Muñoz, D.J.; Bautz, M.W.; Ricker, G.R.; et al. An X-Ray, Infrared, and Submillimeter Flare of Sagittarius A*. {\em Astrophys. J.} {\bf 2008}, {\em 682}, 373.
\bibitem[Do et al.(2019)]{2019ApJ...882L..27D} Do, T.; Witzel, G.; Gautam, A.K.; Chen, Z.; Ghez, A.M.; Morris, M.R.; Becklin, E.E.; Ciurlo, A.; Hosek, M., Jr.; Martinez, G.D.; et al. Unprecedented Near-infrared Brightness and Variability of \sgra. {\em Astrophys. J. Lett.} {\bf 2019}, {\em 882}, L27.
\bibitem[Witzel et al.(2018)]{2018ApJ...863...15W} Witzel, G.; Martinez, G.; Hora, J.; Willner, S.P.; Morris, M.R.; Gammie, C.; Becklin, E.E.; Ashby, M.L.N.; Baganoff, F.; Carey, S.; et al. Variability Timescale and Spectral Index of Sgr A* in the Near Infrared: Approximate Bayesian Computation Analysis of the Variability of the Closest Supermassive Black Hole. {\em Astrophys. J.} {\bf 2018}, {\em 863}, 15.
\bibitem[Eckart et al.(2013)]{2013A&A...551A..18E} Eckart, A.; Mu{\v{z}}i{\'c}, K.; Yazici, S.; Sabha, N.; Shahzamanian, B.; Witzel, G.; Moser, L.; Garcia-Marin, M.; Valencia-S, M.; Jalali, B.; et al. Near-infrared proper motions and spectroscopy of infrared excess sources at the Galactic center. {\em Astron. Astrophys.} {\bf 2013}, {\em 551}, A18.
\bibitem[Do et al.(2019)]{2019Sci...365..664D} Do, T.; Hees, A.; Ghez, A.; Martinez, G.D.; Chu, D.S.; Jia, S.; Sakai, S.; Lu, J.R.; Gautam, A.K.; O’neil, K.K.; et al. Relativistic redshift of the star S0-2 orbiting the Galactic Center supermassive black hole. {\em Science} {\bf 2019}, {\em 365}, 664.
\bibitem[Ressler et al.(2018)]{2018MNRAS.478.3544R} Ressler, S.M.; Quataert, E.; Stone, J.M. Hydrodynamic simulations of the inner accretion flow of Sagittarius A* fuelled by stellar winds. {\em Mon. Not. R. Astron. Soc.} {\bf 2018}, {\em 478}, 3544.
\bibitem[Kawashima et al.(2017)]{2017PASJ...69...43K} Kawashima, T.; Matsumoto, Y.; Matsumoto, R. A possible time-delayed brightening of the Sgr A* accretion flow after the pericenter passage of the G2 cloud. {\em Publ. Astron. Soc. Jpn.} {\bf 2017}, {\em 69}, 43.
\bibitem[Akiyama et al.(2014)]{2014IAUS..303..288A} Akiyama, K.; Kino, M.; Sohn, B.; Lee, S.; Trippe, S.; Honma, M. Long-term monitoring of Sgr A* at 7 mm with VERA and KaVA. {\em  Galact. Center: Feed. Feedback Norm. Galact. Nucl.} {\bf 2014}, {\em 303}, 288.
\bibitem[Lee et al.(2014)]{2014AJ....147...77L} Lee, S.S.; Petrov, L.; Byun, D.Y.; Kim, J.; Jung, T.; Song, M.G.; Oh, C.S.; Roh, D.G.; Je, D.H.; Wi, S.O.; et al. Early Science with the Korean VLBI Network: Evaluation of System Performance. {\em Astron. J.} {\bf 2014}, {\em 147}, 77.
\bibitem[Kino(2015)]{kino15} Kino, M.; Niinuma, K.; Zhao, G.-Y.; Sohn, B.W. Key Science Observations of Agns with the Kava Array. {\em Publ. Korean Astron. Soc.} {\bf 2015}, {\em 30}, 633.
\bibitem[Wajima et al.(2016)]{2016ASPC..502...81W} Wajima, K.; Hagiwara, Y.; An T.; Baan, W.A.; Fujisawa, K.; Hao, L.; Jiang, W.; Jung, T.; Kawaguchi, N.; Kim, J.; et al. The East-Asian VLBI Network. {\em Front. Radio Astron. Fast Early Sci. Symp.} {\bf 2016}, {\em 502}, 81
\bibitem[An et al.(2018)]{2018NatAs...2..118A} An T.; Sohn, B.W.; Imai, H. Capabilities and prospects of the East Asia Very Long Baseline Interferometry Network. {\em Nature Astron.} {\bf 2018}, {\em 2}, 118.
\bibitem[Cui(2021)]{2021RAA....21..205C} Cui, Y.Z.; Hada, K.; Kino, M.; Sohn, B.W.; Park, J.; Ro, H.W.; Sawada-Satoh, S.; Jiang, W.; Cui, L.; Honma, M.; et al. East Asian VLBI Network observations of active galactic nuclei jets: Imaging with KaVA+Tianma+Nanshan. {\em Res. Astron. Astrophys.} {\bf 2021}, {\em21}, 205.
\bibitem[Greisen(2003)]{2003ASSL..285..109G} Greisen, E.W.\ AIPS, the VLA, and the VLBA. {\em Inf. Handl. Astron.-Vistas} {\bf 2003}, {\em 285}, 109.
\bibitem[Lee et al.(2015)]{2015JKAS...48..229L} Lee, S.S.; Byun, D.Y.; Oh, C.S.; Kim, H.R.; Kim, J.; Jung, T.; Oh, S.J.; Roh, D.G.; Jung, D.K.; Yeom, J.H. Amplitude Correction Factors of Korean VLBI Network Observations. {\em J. Korean Astron. Soc.} {\bf 2015}, {\em 48}, 229.
\bibitem[Cho et al.(2017)]{2017PASJ...69...87C} Cho, I.; Jung, T.; Zhao, G.Y.; Akiyama, K.; Sawada-Satoh, S.; Kino, M.; Byun, D.Y.; Sohn, B.W.; Shibata, K.M.; Hirota, T.; et al. A comparative study of amplitude calibrations for the East Asia VLBI Network: A priori and template spectrum methods. {\em Publ. Astron. Soc. Jpn.} {\bf 2017}, {\em 69}, 87.
\bibitem[Shepherd(1997)]{1997ASPC..125...77S} Shepherd, M.C. Difmap: An Interactive Program for Synthesis Imaging. {\em Astron. Data Anal. Softw. Syst. VI} {\bf 1997}, {\em 125}, 77.
\bibitem[Cho et al.(2022)]{2022ApJ...926..108C} Cho, I.; Zhao, G.Y.; Kawashima, T.; Kino, M.; Akiyama, K.; Johnson, M.D.; Issaoun, S.; Moriyama, K.; Cheng, X.; Algaba, J.C.; et al. The Intrinsic Structure of Sagittarius A* at 1.3 cm and 7 mm. {\em Astrophys. J.} {\bf 2022}, {\em 926}, 108.
\bibitem[Lu et al.(2011)]{2011A&A...525A..76L} Lu, R.-S.; Krichbaum, T.P.; Eckart, A.; König, S.; Kunneriath, D.; Witzel, G.; Witzel, A.; Zensus, J.A. Multiwavelength VLBI observations of Sagittarius A*. {\em Astron. Astrophys.} {\bf 2011}, {\em 525}, A76.
\bibitem[Johnson et al.(2018)]{2018ApJ...865..104J} Johnson, M.D.; Narayan, R.; Psaltis, D.; Blackburn, L.; Kovalev, Y.Y.; Gwinn, C.R.; Zhao, G.Y.; Bower, G.C.; Moran, J.M.; Kino, M.; et al. The Scattering and Intrinsic Structure of Sagittarius A* at Radio Wavelengths. {\em Astrophys. J.} {\bf 2018}, {\em 865}, 104.
\bibitem[Akiyama et al.(2013)]{2013PASJ...65...91A} Akiyama, K.; Takahashi, R.; Honma, M.; Oyama, T.; Kobayashi, H. Multi-Epoch VERA Observations of Sagittarius A*. I. Images and Structural Variability. {\em Publ. Astron. Soc. Jpn.} {\bf 2013}, {\em 65}, 91.
\bibitem[Jorstad et al.(2017)]{Jorstad_2017} Jorstad, S.G.; Marscher, A.P.; Morozova, D.A.; Troitsky, I.S.; Agudo, I.; Casadio, C.; Foord, A.; Gómez, J.L.; MacDonald, N.R.; Molina, S.N.; et al. Kinematics of Parsec-scale Jets of Gamma-Ray Blazars at 43 GHz within the VLBA-BU-BLAZAR Program. {\em Astrophys. J.} {\bf 2017}, \emph{846}, 98. https://doi.org/10.3847/1538-4357/aa8407.
\bibitem[Weaver et al.(2022)]{Weaver_2022} Weaver, Z.R.; Jorstad, S.G.; Marscher, A.P.; Morozova, D.A.; Troitsky, I.S.; Agudo, I.; Gómez, J.L.; Lähteenmäki, A.; Tammi, J.; Tornikoski, M. Kinematics of Parsec-scale Jets of Gamma-Ray Blazars at 43 GHz during 10 yr of the VLBA-BU-BLAZAR Program. {\em Astrophys. J. Supplement} {\bf 2022}, \emph{260}, 12. https://doi.org/10.3847/1538-4365/ac589c.
\bibitem[Lee et al.(2016)]{Lee_2016} Lee, S.S.; Wajima, K.; Algaba, J.C.; Zhao, G.Y.; Hodgson, J.A.; Kim, D.W.; Park, J.; Kim, J.Y.; Miyazaki, A.; Byun, D.Y.; et al. Interferometric Monitoring of Gamma-Ray Bright AGNs. I. The Results of Single-epoch Multifrequency Observations. 
{\em Astrophys. J. Supplement} {\bf 2016}, \emph{227}, 8. https://doi.org/10.3847/0067-0049/227/1/8.
\bibitem[Bower et al.(2014)]{2014ApJ...790....1B} Bower, G.C.; Markoff, S.; Brunthaler, A.; Law, C.; Falcke, H.; Maitra, D.; Clavel, M.; Goldwurm, A.; Morris, M.R.; Witzel, G.; et al. The Intrinsic Two-dimensional Size of Sagittarius A*. {\em Astrophys. J.} {\bf 2014}, {\em 790}, 1.
\bibitem[Markoff et al.(2007)]{2007MNRAS.379.1519M} Markoff, S.; Bower, G.C.; Falcke, H.. How to hide large-scale outflows: Size constraints on the jets of \sgra. {\em Mon. Not. R. Astron. Soc.} {\bf 2007}, {\em 379}, 1519.
\bibitem[Psaltis et al.(2018)]{2018arXiv180501242P} Psaltis, D.; Johnson, M.; Narayan, R.; Medeiros, L.; Blackburn, L.; Bower, G. A Model for Anisotropic Interstellar Scattering and its Application to \sgra. \emph{arXiv} {\bf 2018},  arXiv:1805.01242.
\bibitem[Curran(2014)]{curran_2014} Curran, P.A. Monte Carlo error analyses of Spearman's rank test. \emph{arXiv} \textbf{2014}, arXiv:1411.3816.
\bibitem[Wielgus et al.(2022)]{2022A&A...665L...6W} Wielgus, M.; Moscibrodzka, M.; Vos, J.; Gelles, Z.; Martí-Vidal, I.; Farah, J.; Marchili, N.; Goddi, C.; Messias, H. Orbital motion near Sagittarius A*. Constraints from polarimetric ALMA observations. {\em Astron. Astrophys.} {\bf 2022}, {\em 665}, L6.
\bibitem[Narayan et al.(1995)]{1995Natur.374..623N} Narayan, R.; Yi, I.; Mahadevan, R. Explaining the spectrum of Sagittarius A* with a model of an accreting black hole. {\em Nature} {\bf 1995}, {\em 374}, 623.
\bibitem[{\"O}zel et al.(2000)]{2000ApJ...541..234O} {\"O}zel, F.; Psaltis, D.; Narayan, R. Hybrid Thermal-Nonthermal Synchrotron Emission from Hot Accretion Flows. {\em Astrophys. J.} {\bf 2000}, {\em 541}, 234.
\bibitem[Shen et al.(2005)]{2005Natur.438...62S} Shen, Z.-Q.; Lo, K.Y.; Liang, M.-C.; Ho, P.T.P.; Zhao, J.-H. A size of 1au for the radio source \sgra at the centre of the Milky Way. {\em Nature} {\bf 2005}, {\em 438}, 62.
\bibitem[Alberdi et al.(1993)]{1993A&A...277L...1A} Alberdi, A.; Lara, L.; Marcaide, J.M.; Elosegui, P.; Shapiro, I.I.; Cotton, W.D.; Diamond, P.J.; Rommey, J.D.; Preston, R.A. VLBA image of SGR A at lambda = 1.35 cm. {\em Astron. Astrophys.} {\bf 1993}, {\em 277}, L1.
\bibitem[Marcaide et al.(1999)]{1999A&A...343..801M} Marcaide, J.M.; Alberdi, A.; Lara, L.; Pérez-Torres, M.A.; Diamond, P.J. A decade of unchanged 1.3 CM VLBI structure of SGR A*. {\em Astron. Astrophys.} {\bf 1999}, {\em 343}, 801.
\bibitem[Ressler et al.(2020)]{2020ApJ...896L...6R} Ressler, S.M.; White, C.J.; Quataert, E.; Stone, J.M. Ab~Initio Horizon-scale Simulations of Magnetically Arrested Accretion in Sagittarius A* Fed by Stellar Winds. {\em Astrophys. J. Lett.} {\bf 2020}, {\em 896}, L6.
\bibitem[Kawashima et al.(2019)]{2019ApJ...878...27K} Kawashima, T.; Kino, M.; Akiyama, K. Black Hole Spin Signature in the Black Hole Shadow of M87 in the Flaring State. {\em Astrophys. J.} {\bf 2019}, {\em 878}, 27.
\bibitem[Kawashima et al.(2021)]{arXiv:2108.05131} Kawashima, T.; Ohsuga, K.; Takahashi, H.R. RAIKOU: A General Relativistic, Multi-wavelength Radiative Transfer Code. \emph{arXiv} {\bf 2021}, arXiv:2108.05131.
\bibitem[Rauch et al.(2016)]{Rauch_2016} Rauch, C.; Ros, E.; Krichbaum, T.P.; Eckart, A.; Zensus, J.A.; Shahzamanian, B.; Mužić, K. 
Wisps in the Galactic center: Near-infrared triggered observations of the radio source Sgr A* at 43 GHz. 
{\em Astron. Astrophys.} {\bf 2016}, 587, A37. \url{https://doi.org/10.1051/0004-6361/201527286}.

\end{thebibliography}
\end{document}